\documentclass[]{aa}
\usepackage{graphicx}
\usepackage{txfonts}
\newcommand{\epsfig}[1]{\resizebox{\hsize}{!}{\includegraphics{#1}}}
\newcommand{\figcap}[1]{\caption[Another figure]{#1}}
\newcommand{\AJ}[3]{{#1}, AJ, \vol{{#2}}, {#3}.}
\newcommand{\ApJ}[3]{{#1}, ApJ, \vol{{#2}}, {#3}.}
\newcommand{\ApJS}[3]{{#1}, ApJS, \vol{{#2}}, {#3}.}

\newcommand{\AandA}[3]{{#1}, A\&A, \vol{{#2}}, {#3}.}
\newcommand{\AandAsupp}[3]{{#1}, A\&A Supp.\rm, \vol{{#2}}, {#3}.}

\newcommand{\MNRAS}[3]{{#1}, MNRAS\rm, \vol{{#2}}, {#3}.}

\newcommand{\PASP}[3]{{#1}, PASP\rm, \vol{{#2}}, {#3}.}

\newcommand{\vol}[1]{{\mbox{#1}}}

\newcommand{\Teff}{\mbox{$T_{\mbox{\scriptsize eff}}\,$}}

\newcommand{\FeH}{\mbox{[Fe/H]}\,}

\newcommand{\kms}{\mbox{$\mbox{km\,s}^{-1}$}\,}

\newcommand{\logg}{\mbox{$\log g$}\,}

\newcommand{\logP}{\mbox{$\log P$}\ }

\newcommand{\magdex}{\mbox{mag~dex$^{-1}\,$}}

\input{epsf.tex}
\begin{document}
\title{The effect of metallicity on the Cepheid Period-Luminosity 
       relation from a
       Baade-Wesselink analysis of Cepheids in the Galaxy and in the
       Small Magellanic Cloud
         \thanks{Some of the observations reported
          here were obtained with the Multiple Mirror Telescope, operated
          jointly by the Smithsonian Institution and the University of
          Arizona.
         }
	 $\!^{,}\!$
         \thanks{
               {    Tables A.2-A.11 are only available in
                    electronic form at the CDS via anonymous ftp
                    to cdsarc.u-strasbg.fr or via
                    http://cdsweb.u-strasbg.fr/A+A.htx
               }
         }
       }


 \author{Jesper~Storm\inst{1}
 \and 
	 Bruce~W.~Carney\inst{2}
 \and
         Wolfgang~P.~Gieren\inst{3}
 \and
         Pascal~Fouqu\'e\inst{4,5}
 \and
	 David~W.~Latham\inst{6}
 \and
         Anne~M.~Fry\inst{2}
}
%

\offprints{J. Storm, \email{jstorm@aip.de}}

 \institute{Astrophysikalisches Institut Potsdam,
            An der Sternwarte 16, D-14482 Potsdam, Germany,
	e-mail: jstorm@aip.de
 \and
            Univ. of North Carolina at Chapel Hill,
            Dept. of Physics and Astronomy,
            Phillips Hall, Chapel Hill,
            NC-27599-3255, USA\\
	e-mail: bruce@physics.unc.edu, anne.fry@integraonline.com
 \and
           Universidad de Concepci\'on, Departamento de F\'{\i}sica,
	   Casilla 160-C,
           Concepci\'on, Chile,
	e-mail: wgieren@coma.cfm.udec.cl
\and
           Observatoire de Paris, LESIA,
           5, Place Jules Janssen,
           F-92195 Meudon Cedex, France
 \and
	   European Southern Observatory,
           Casilla 19001,
           Santiago 19, Chile
	e-mail: pfouque@eso.org
\and
           Harvard-Smithsonian Center for Astrophysics,
           60 Garden Street, Cambridge,
           Massachusetts 02138, USA, e-mail: dlatham@cfa.harvard.edu
}

\date{Received 5 August 2003 / Accepted 14 November 2003}

\titlerunning{The metallicity effect on the Cepheid P-L relation}

\abstract{
We have applied the near-IR Barnes-Evans realization of the
Baade-Wesselink method as calibrated by Fouqu\'e \& Gieren (\cite{FG97})
to five metal-poor Cepheids with periods between 13 and 17 days
in the Small Magellanic Cloud as well as to
a sample of 34 Galactic Cepheids to determine the effect of metallicity
on the period-luminosity (P-L) relation. For ten of the Galactic Cepheids
we present new accurate and well sampled radial-velocity curves.
The Baade-Wesselink analysis provides accurate individual distances
and luminosities to the Cepheids in the two samples, allowing us to
constrain directly, in a purely differential way, the metallicity effect
on the Cepheid P-L relation.  For the Galactic Cepheids we provide a new
set of P-L relations which have zero-points in excellent agreement with
astrometric and interferometric determinations. These relations can be
used directly for the determination of distances to solar-metallicity
samples of Cepheids in distant galaxies, circumventing any corrections
for metallicity effects on the zero-point and slope of the P-L relation.
We find evidence for both such metallicity effects in our data.
Comparing our two samples of Cepheids at a mean period of about 15~days,
we find a weak effect of metallicity on the luminosity similar
to that adopted by the HST
Key Project on the Extragalactic Distance Scale. The effect is smaller
for the $V$~band, where we find $\Delta M_V / \Delta \FeH =
-0.21\pm0.19$, and larger for the Wesenheit index $W$, where we
find $\Delta M_W / \Delta \FeH = -0.29\pm0.19$. For the $I$ and $K$~bands
we find $\Delta M_I / \Delta \FeH = -0.23\pm
0.19$ and $\Delta M_K / \Delta \FeH = -0.21\pm 0.19$, respectively.
The error estimates are 1~$\sigma$ statistical errors.
It seems now well established that metal-poor Cepheids with periods
longer than about 10 days are intrinsically fainter in all these
bands than their metal-rich counterparts of identical period.
Correcting the LMC distance estimate of Fouqu\'e et al. (\cite{FSG03})
for this metallicity effect leads to a revised LMC distance modulus of 
$(m-M)_0 = 18.48\pm 0.07$, which is also in excellent agreement with
the value of $(m-M)_0 = 18.50\pm 0.10$ adopted by the Key Project.
From our SMC Cepheid distances we determine the \object{SMC} distance
to be $18.88\pm0.13$~mag irrespective of metallicity. 

}

\maketitle

\section{Introduction}

  Cepheids remain one of the most accurate and well understood
standard candles providing distances to galaxies out to the Virgo
cluster. The Hubble Space Telescope has been used extensively to
determine distances to such distant galaxies through the application of
the Cepheid period-luminosity (P-L) relation. In this way it has been
possible to calibrate secondary distance indicators which in turn
are applicable to cosmological distances thus enabling an accurate
determination of the Hubble constant. The most recent papers in these
series are Saha et al. (\cite{Saha01}), who calibrate the peak luminosity
of type Ia supernovae and find a Hubble constant of 
$H_0 = 58.7 \pm 6.3 \mbox{(internal)} \kms \mbox{Mpc}^{-1}$ and
Freedman et al. \cite{Freedman01}, who presents the final results of the
Hubble Space Telescope Key Project on the Extragalactic Distance
Scale, and find a value of $H_0 = 72 \pm 8 \kms \mbox{Mpc}^{-1}$.

  These results are based on a few qualified assumptions. The first
assumption is that the distance modulus to the Large Magellanic Cloud (LMC) is
$\mu_{0} = 18.50\pm0.10$, and the second assumption is that the P-L relation is
only weakly dependent on metallicity. Freedman et al.  (\cite{Freedman01})
adopt the value 
$\gamma_{VI} = \delta \mu_0 / \delta \FeH = -0.2\pm0.2$~mag dex$^{-1}$
based on the investigation of Kennicutt et al. (\cite{Kenn98}).

  The first assumption has been the subject of extensive studies, but there is
still not a complete consensus about the distance to this galaxy, which
is one of our nearest neighbours.  Recent distance estimates to the
LMC still range, embarrasingly enough, from 18.1 to 18.8, 
see e.g. Benedict et al.  (\cite{Benedict02}).

  The metallicity effect has not been investigated quite as thoroughly
and from both the observational and theoretical points of view the results
have not been very accurate. This means that there is even doubt about
the sign of the possible effect, and it is also not entirely clear if 
the slope of the P-L relation (Tammann et al. \cite{Tammann03}) is affected.

  The Baade-Wesselink method and its variants provide the means to
determine accurate distances to {\em individual} Cepheids in contrast to
the P-L relation which, due to the finite width of the instability strip,
only provides accurate determinations to an ensemble of Cepheids.
Several calibrations of the Baade-Wesselink method based on interferometric
diameters measured for red giant stars have appeared in the literature
over the last decade, see e.g. Fouqu\'e \& Gieren (\cite{FG97}, FG97
hereinafter), Di
Benedetto (\cite{DiBenedetto97}), or Arellano Ferro \& Rosenzweig
(\cite{Arellano00}), and they show great promise for
delivering accurate distances only weakly dependent on stellar
atmosphere computations (Gieren et al. \cite{Gieren00}). These
calibrations are currently being confirmed by direct interferometric
measurements of nearby Cepheids (Nordgren et al.
\cite{Nordgren2002}).

  In the present work we investigate the difference in absolute
magnitude between Cepheids with different metallicity. 
We do this by comparing absolute magnitudes for five metal-poor
Cepheids in the Small Magellanic Cloud ($\FeH=-0.7$) with 
solar-metallicity Cepheids in the Milky Way. The SMC Cepheids are the most
metal-poor Cepheids which we have access to and provide a lever-arm in
metallicity which is twice as large as for Cepheids in the LMC.
The Galactic Cepheids are similarly the most metal-rich Cepheids
which are sufficiently close to allow
detailed observations. In both cases we determine the absolute magnitudes
by applying the Infrared Surface-Brightness (hereinafter ISB) method
calibrated by FG97. This method is a near-IR calibration
of the Barnes-Evans (Barnes \& Evans \cite{BarnesEvans76}) variant of
the Baade-Wesselink method.

By applying the same method to both samples of stars we can
determine, in a purely {\em differential} way, the effect of metallicity,
independent of the exact luminosity zero-point of the method.

  In Sect.2 we present the available observational data for the SMC
sample of five Cepheids as well as for the sample of Galactic Cepheids
which we will use for the comparison. We also discuss the reddening
and reddening law for the two samples which is crucial for the final
conclusions. New radial-velocity curves for ten of the Galactic Cepheids
can be found in the Appendix.

In Sect.3 we present the FG97
$F_V,(V-K)$-version of the Barnes-Evans surface brightness method which we
use, and we argue that the method itself is
very insensitive to reddening and metallicity effects.

In Sect.4 we present our results as they
emerge from our application of the ISB method to the two samples of
stars. We determine new Galactic P-L relations in the $BVIJHKW$~bands
based on the ISB analysis of the Galactic sample of stars. We compare
the luminosities for the SMC stars to these relations and find a weak
metallicity effect. We proceed to apply this metallicity effect to the
LMC Cepheids and find a new metallicity corrected distance modulus for the LMC.

In Sect.5 we discuss the results in context with other determinations and
we show that both the metallicity effect and the metallicity corrected
LMC distance modulus are in excellent agreement with the values adopted by the Key
Project (Freedman et al. \cite{Freedman01}).

\section{Observational Parameters}

\begin{table}
\caption{\label{tab.GCref}The list of the references which provided the 
$V$ and $K$~band photometry and the radial velocities.}
\begin{tabular}{r c c c}
\hline\hline
Star & $V$ & $K$ & Radial velocity \\
\hline
$\eta$ Aql	&3,14&3,20&6,10,12\\
WZ Car	&5,17&11,19&5\\
U Car	&5,17&11,20&5,12\\
l Car	&17&11&12\\
VY Car	&5,13,17&11,20&5,12\\
SU Cas	&3,14&3&2,6\\
KN Cen	&5&11&5\\
XX Cen	&5&11&12\\
VW Cen	&5,17&11&5\\
V Cen	&9,17&11,20&8,12\\
        &   &   &  \\
$\delta$ Cep	&3,14&3&6,10\\
X Cyg	&10,14&3,20&2,6\\
T Mon	&14&11&2,6\\
CV Mon	&14,17&11,20&6,16\\
UU Mus	&5,13&11&5\\
U Nor	&5,13,17&11&5,18\\
S Nor	&17&11&2\\
V340 Nor	&22,7&11&2\\
BF Oph	&9,14&11&1,8,12\\
LS Pup	&5&11,19&5\\
        &   &   &  \\
BN Pup	&5,13,17&11,19&5\\
VZ Pup	&5,17&11,19&5\\
AQ Pup	&5,14&11,19,21&1,5\\
RS Pup	&17&11,20,21&6\\
RY Sco	&5,13,14,17&11,20&1,5,12\\
EV Sct	&14,17&11&2,6,15\\
BB Sgr	&9,14&11,20&8\\
WZ Sgr	&14&11,20&1,5\\
U Sgr	&9,14&11&2,6\\
T Vel	&17,23&11&8,12\\
        &   &   &  \\
SW Vel	&13,14,17&11&5,18\\
RY Vel	&5&11,20&5\\
RZ Vel	&5,13&11&5,12\\
SV Vul	&14&3,11,20&6\\
\hline
\end{tabular}


{\scriptsize
1: Barnes et al., \cite{BMS87};
2: Bersier et al. \cite{Bersier94};
3: Barnes, et al. \cite{Barnes97};
4: Berdnikov, \& Turner, \cite{BT95};
5: Coulson, \& Caldwell, \cite{Coulson85a};
6: This work;
7: Eggen, \cite{Eggen83};
8: Gieren, \cite{Gieren81a};
9: Gieren, \cite{Gieren81b};
10: Kiss, \cite{Kiss98};
11: Laney, \& Stobie, \cite{LS92};
12: Lloyd Evans, \cite{Lloyd80};
13: Madore, \cite{Madore75};
14: Moffett, \& Barnes \cite{MB84};
15: Metzger et al. \cite{Metzger91};
16: Metzger et al., \cite{Metzger92};
17: Pel, \cite{Pel76};
18: Pont, et al., \cite{Pont94};
19: Schechter et al., \cite{SACK92};
20: Welch, et al., \cite{Welch84};
21: Welch, \cite{Welch85};
22: Coulson, \& Caldwell, \cite{Coulson85b}
23: Gieren, \cite{Gieren85};
}
\end{table}

\subsection{The \object{SMC} Cepheid photometry and radial velocities}

  The sample of five \object{SMC} Cepheids and their $BVRIJK$ photometry
and radial-velocity curves have been presented in full in Storm et al.
(\cite{Storm03}). Here we emphasize that the five \object{SMC} Cepheids
are fundamental mode pulsators with periods of approximately 15~days.
With a sample of five stars at similar period, we have a good
constraint on the location of the ridge line of the P-L relation, at this period.

  The optical light curves are
based on CCD photometry presented in that paper as well as data from
Udalski et al. (\cite{Udalski99b}), and the near-IR light curves are based
on data obtained with various near-IR imaging cameras at the Las Campanas,
Cerro Tololo and La Silla observatories.

  The radial-velocity curves are based on high resolution
echelle spectra obtained mainly at the Las Campanas Observatory
but supplemented with data from both Cerro Tololo and La Silla.

  The $(V-K)$ color curve which is used for the Baade-Wesselink analysis
has been derived for each Cepheid by performing a low order Fourier fit to
the mostly sinusoidal and low amplitude $K$~band light curve and
determining the $K$-magnitude from the fit at every $V$~band
observation as described in Storm et al. (\cite{Storm03}). This procedure
was necessary as the optical and near-IR data were not obtained
simultaneously.

\subsection{The Milky Way Cepheid photometry and radial velocities}

  The reference sample of fundamental mode Galactic Cepheids was selected
on the availability of accurate $V$ and $K$ light curves as well as
accurate radial
velocity curves. For the present purpose it is important both
that the data points have small uncertainties and also that the phase
coverage is good in all three observables. The data for these stars were
compiled from the literature and by using the McMaster Cepheid
database. In addition we have acquired extensive radial-velocity data
for a number of these stars. These data are presented in the Appendix.
The original sources for the data for each star can be found in
Table \ref{tab.GCref}.

  We have also compiled the $B$ and $I$ data from these sources and have
transformed the $I$ data to the Cousins system using the transformations
from the Johnson system given by Caldwell et al. (\cite{CCG85})
where necessary. We have finally also included the intensity mean $I$
values collected by Groenewegen (\cite{Groen99}) for S~Nor, l~Car, and RS~Pup.

\subsection{The reddening towards the Galactic Cepheids}
\label{sec.GalRed}

  Fernie (\cite{Fernie90}) has combined the available reddening estimates
toward Galactic Cepheids from many different photometric systems
in a systematic way.  He has furthermore determined a transformation from the
resulting reddening scale to the cluster reddening scale of Feast \& Walker
(\cite{FW87}). The latter scale is based on cluster OB star reddenings.
We adopt these transformed reddenings so we are sure that we are on the
same system as used for the ZAMS fitting distances to the clusters and
thus that we can compare our absolute $V$ magnitudes with those found from
ZAMS fitting to the clusters.

  We have adopted the optical reddening laws from Caldwell \& Coulson
(\cite{CC87}) and Dean et al. (\cite{DWC78}) and the near-IR
reddening laws from Laney \& Stobie (\cite{LS93}).

  We define the coefficient $R_x$ in the band $x$ from the absorption in
the band $A_x$ by $A_x=R_x \times E(B-V)$ or
$R_x = A_x / E(B-V) = R_V \times A_x / A_V$.

We thus adopt the following relations:
\begin{eqnarray}
R_V & = & 3.07 + 0.28(B-V)_0 + 0.04E(B-V)\\
R_I & = & 1.82 + 0.205(B-V)_0 + 0.0225E(B-V)\\
R_J & = & 0.249 \times R_V\\
R_H & = & 0.147 \times R_V\\
R_K & = & 0.091 \times R_V
\end{eqnarray}

  For the reddening insensitive Wesenheit function $W_{VI}$ 
(Freedman et al.  \cite{Freedman01}) we have $W = V - R_W (V - I)$
where $R_W = 1 / (1-(R_I/R_V))$.

  Computing the mean over the 32 Galactic calibrators we find that the
dispersion in $R_x$ is very small (0.036 in $V$) and we simply adopt constant
values for $R_x$. We find $R_V=3.30$, $R_I=1.99$, $R_W=2.51$,
$R_J=0.82$, $R_H=0.48$, and $R_K=0.30$.

We note that following Cardelli et al. (\cite{Cardelli89}) $R_W$
would equal 2.45, but we prefer the above value as it is based directly
on Cepheid measurements.

\subsection{The adopted metallicity}
\label{sec.SMC_FeH}

  It is well known that the Galactic Cepheids exhibit a range of
metallicity values (see e.g. Fry \& Carney \cite{Fry97} and Luck et al.
\cite{Luck03}), and it is
expected that the \object{SMC} Cepheids similarly span a range of values.
However, as the \object{SMC} stars are faint from a high-resolution
spectroscopic point of view, we do not have individual spectroscopic
abundances for these stars yet. Instead we adopt an average value for
our sample of stars as we are here mainly interested in the
relative differences between the Milky Way Cepheids and the \object{SMC}
Cepheids.

  Luck et al. (\cite{Luck98}) determined
metallicities for six \object{SMC} Cepheids based on high-resolution
spectroscopy. Combined with similar data for an additional 19 Cepheids
from the literature these authors find the average value $\FeH=-0.68$
with a standard deviation of $\sigma=0.13$.  The same authors find
from a similar study
of Galactic Cepheids an average value of $\FeH=-0.03$ based on their own
sample of 11 stars, and an average value of $\FeH=+0.03$ ($\sigma=0.14$)
when averaged over an additional 58 stars from the literature. Combining
these two numbers gives a metallicity offset of
$\Delta \FeH = \FeH(\mbox{SMC}) - \FeH(\mbox{MW}) = -0.70$
which we will adopt in the following. It is clear that the intrinsic
spread introduces an additional uncertainty on the final result and
individual metallicities for the stars would be preferable. The spread
for the Galactic Cepheids should average out as the Baade-Wesselink
analysis has been extended to a large number of stars, but for the five
\object{SMC} stars we estimate an additional uncertainty of the mean of
$\sigma = 0.13 / \sqrt{5-1} = 0.07$dex for the adopted mean metallicity.

\subsection{The reddening towards the \object{SMC} Cepheids}

Bessell (\cite{Bessell91}) discusses in detail a large body of reddening
investigations for the SMC and finds that the foreground reddening
towards the SMC is smooth with a mean value of about $E(B-V)=0.05$. He
argues that the average reddening within the SMC is of a similar size
but notes that different methods and populations give different answers.

  Zaritsky et al. (\cite{Zar02}) find a mean absorption of
$A_V=0.18$~mag ($E(B-V)=0.055$ for $R_V=3.30$) for their sample of cool
stars ($5500 \mbox{K} \le T_{\mbox{\scriptsize eff}} \le 6500$~K) with
very little spatial variation, but they also find a significantly higher
mean absorption ($A_V=0.46$) for their sample of hot stars.

Larsen et al. (\cite{Larsen2000}) determined
reddenings towards SMC B-stars using Str\"omgren photometry. They
found foreground reddenings of $E(B-V)=0.07\pm0.02$ and mean reddenings
of $E(B-V)=0.13$ for the two fields they investigated, in good agreement
with the hot population results from Zaritsky et al. (\cite{Zar02})
but larger than the foreground value of $E(B-V)=0.037$
given by Schlegel et al. \cite{Schlegel98}. The latter result is
based on DIRBE measurements of foreground dust emission around the SMC.
Larsen et al.  (\cite{Larsen2000}) also found that the reddening within
each field exhibits significant scatter even on very small spatial
scales (few arcsec) 
suggesting that reddening internal to the SMC and depth effects contribute
significantly to the reddening of the individual stars.

Caldwell \& Coulson (\cite{CC85}) determined $E(B-V)=0.07$ for
\object{HV1335} using $BVI$ reddenings based on the method of Dean et
al. (\cite{DWC78}) recalibrated for the lower metallicity of the
SMC. Laney \& Stobie (\cite{LS94}) have adopted this value together
with a value of $E(B-V)=0.06$ towards the other four stars in our sample
based on the mean reddening calibrated with the Caldwell \& Coulson 
(\cite{CC85}) relation. These values have been tabulated in
Table \ref{tab.BVIred}.  In this table we have also 
tabulated the reddening values assigned by OGLE-2 (Udalski et al.
\cite{Udalski99b}) to the stars in common. These
reddenings are based on red giant clump star colours and averaged on a
field by field basis and they are only slightly
larger than the values from Laney \& Stobie (\cite{LS94}).

The apparent complication of determining accurate reddenings towards SMC
objects using a reddening map type approach shows that direct
reddening determinations for individual objects are necessary for
obtaining accurate luminosities, especially in the reddening sensitive
bands.  As a consequence we have derived direct
measures for the program stars using the observed mean $BVI$ data from Storm
et al.  (\cite{Storm03}) and a recalibrated version of the Dean et al.
(\cite{DWC78}) method.

The Dean et al. (\cite{DWC78}) method builds up an intrinsic
(non-reddened) locus in a $(B-V)$ vs. $(V-I)$ diagram based on $BVI$
photometry of individual Cepheids. For each star the different phase
points fall very narrowly along a straight line in the $(B-V), \  (V-I)$
plane. The straight lines for the different stars are shifted along the
reddening vector to coincide and build up the intrinsic locus.
The zero point is based on Cepheids in clusters with independent reddening
estimates, typically based on B-stars. The locus of the intrinsic relation
is sensitive to metallicity and we have corrected this effect using
model atmosphere computations. For this purpose Daniel Cordier has kindly
evolved an 8 solar-mass model through the second and third crossing of
the blue loop for three metallicities,
$Z=0.020$ (solar), $Z=0.008$ (LMC), and $Z=0.004$ (SMC) using his model 
atmospheres and the BaSeL library (Lejeune et al.
\cite{Lejeune98}). The paths of the second and third
crossings are similar in the color-color diagram and can be well represented
by a straight line.
For each metallicity we have fitted a linear $(B-V)$ vs. $(V-I)$
relation, and we find a constant slope with metallicity but a
blue shift in the intercept of $-0.054$ (LMC) and $-0.096$ (SMC) for fixed
$(V-I)$, which corresponds to a shift in $E(B-V)$ of $-0.08$ and $-0.14$
for the LMC and SMC, respectively. These results are in excellent
agreement with the original Dean et al. (\cite{DWC78}) calibration.
The main assumption underlying the method is that unreddened Cepheids
indeed follow a common relation in the $(B-V),(V-I)$ plane independent of
other parameters, luminosity (and thus period) in particular. 

Due to the low absolute value of the reddening the
exact choice of the reddening law does not cause a significant difference
in the dereddened magnitudes. We assume that the Galactic and SMC reddening
laws are quite similar and simply adopt the previously discussed
Galactic law for the SMC Cepheids.

Applying these relations to the observed $BVI$ data we find reddening values
ranging from $-0.02$ to $0.09$ as tabulated in Table \ref{tab.BVIred}.
The negative reddening for HV1328 is clearly unphysical and a value of
0.0 has been adopted for this star in the following. The mean value for
the remaining stars is $0.055$ in good agreement with the Laney \&
Stobie (\cite{LS94}) values, but several of the values are quite low
compared to the available foreground reddening estimates.

 Assuming an uncertainty of
0.01~mag for both the $(B-V)$ and $(V-I)$ colors gives an estimated
uncertainty of about 0.02~mag on the individual estimates to which a similar
amount of systematic uncertainty on the color zero points has to be
added.  A small global offset may still be present, but the 
reddening differences should be real, assuming that all our stars share the 
same metallicity. 

\begin{table}
\caption{\label{tab.BVIred} Reddenings values for the stars from different 
authors}
\begin{tabular}{r c c c}
\hline\hline
\multicolumn{1}{c}{ID} & \multicolumn{3}{c}{$E(B-V)$}\\
       & LS94$^1$ & OGLE-2$^2$ & $BVI^3$ \\
       & mag  & mag & mag \\
\hline
 HV822 & 0.06 & 0.078 & $+0.03$ \\
HV1328 & 0.06 & --    & $-0.02^4$ \\
HV1333 & 0.06 & --    & $+0.07$ \\
HV1335 & 0.07 & 0.070 & $+0.09$ \\
HV1345 & 0.06 & 0.078 & $+0.03$ \\
\hline
\end{tabular}

{\scriptsize
1: Laney \& Stobie \cite{LS94}; 2: Udalski et al. \cite{Udalski99b}; 3:
Using the $BVI$ calibration described in the text; 4:
This value is unphysical and a value of zero has been adopted.}
\end{table}

For the reddening sensitive bands the use of the individual reddenings
results in a slightly smaller scatter for the derived P-L relations.

  We conclude that individual reddening determinations are
still rather uncertain, but as there appears to be a significant intrinsic
scatter in the SMC reddenings, we prefer these values over the average
values available. We note that the zero-point might be slightly too low,
but that we do not have an obviously better estimate available. We
estimate an uncertainty on the $E(B-V)$ zero point of 0.03~mag.

\section{The Baade-Wesselink Analysis}

\subsection{The infrared surface brightness-color relation}

Fouqu\'e \& Gieren (\cite{FG97}) have recalibrated the Barnes-Evans
surface-brightness relation (Barnes \& Evans \cite{BarnesEvans76})
using interferometrically determined radii
for giants and supergiants. They have also extended the calibration to
use not only the $(V-R)$ but also the $(V-K)$ and $(J-K)$ color indices.
Combining the stellar angular diameters determined from the
surface-brightness variation with the radius variation measured from
radial-velocity curves they determine stellar radii and distances.
In the following we will apply this method to our sample of galactic
and SMC Cepheids.

Gieren et al. (\cite{GFG97}) (hereinafter GFG97) applied
the FG97 calibration to a sample of Cepheids belonging to Galactic open
clusters. In this way they could compare the distance estimates from
the FG97 method with independent estimates based on ZAMS fitting. They
found excellent agreement ($5\%$ in the distance) for the $F_V,(V-K)$ and
$F_K,(J-K)$ calibrations whereas the distances based on the $F_V,(V-R)$
calibration showed significant systematic and random errors. This supports
the finding of Welch (\cite{Welch94}) who argued strongly in favour of
using near-IR calibrations of the surface brightness.

  New interferometers have recently started to produce direct measurements
of Cepheid mean radii (Kervella et al. \cite{Kervella2001}), and Cepheid radius
variations (Lane et al. \cite{Lane2002}). Nordgren et al. (\cite{Nordgren2002})
present 59 interferometric measurements for three nearby Cepheids and
find a surface brightness-color calibration which is very similar (to
within 4\% in distance) to that of FG97. The direct interferometric
observations of Cepheids show great promise for placing the
calibration on an even firmer basis in the near future as the
measurements are done directly on pulsating stars of the same color as
the stars to be studied. 
For now we choose to use the FG97 calibration (Eq.(\ref{eq.colFv}))
as it is based on many more stars and
on individual angular diameters which exhibit less scatter.

\subsection{The effect of metallicity and gravity}
\label{sec.ISB_metal}

  The FG97 calibration is based on Galactic stars which all have about
solar abundance but span a large range of surface gravities.
A certain dependence on metallicity is to be expected as
the surface brightness, which is a temperature measure, is calibrated as
a function of a color index.

  To investigate the significance of such an effect
we have taken the FG97 calibration (Eq.(\ref{eq.colFv})) and rearranged
it to include the effective temperature (Eq.(\ref{eq.FvT})).
\begin{eqnarray}
\label{eq.colFv}
F_V & = & 3.947 - 0.131(V-K)_0 \\
    & = & 4.2207 - 0.1 \times S_V \\
\label{eq.FvT}
    & = & 4.2207 - 0.1(42.207 - 10 \log T_{\mbox{\scriptsize eff}} -
B.C._V)
\end{eqnarray}

Using the Kurucz model atmospheres (ATLAS9, Kurucz \cite{Kurucz93})
we have determined a
grid of colors, temperatures, and bolometric corrections for an assumed
value of $\logg = 1.5$ and for a range of different metallicities. For
each metallicity we can then compute $F_V$ for a range of temperatures
and as we have corresponding values of $(V-K)_0$ we can re-derive the linear
relation between $F_V$ and $(V-K)_0$ for each metallicity. The relation
does not reproduce the empirical relation above exactly but gives
instead $F_V=3.949(\pm0.005)-0.126(\pm0.002)(V-K)_0$ for an assumed
solar metallicity. For an assumed metallicity of $\FeH=-0.7$ the relation
becomes $F_V=3.947-0.125(V-K)_0$, which means that there is no significant
difference between the two relations. 

  We have repeated the procedure for \logg
= 0.75 and \logg = 2.25 thus covering a range in \logg
similar to that spanned by the Cepheids in question according to the
relation $\logg = 2.620 - 1.142 \log P$ from Fernie (\cite{Fernie95}).
Again the differences are small. For the star X~Cyg we find
a distance estimate which is larger by 0.03~mag for the \logg=0.75
relation and smaller by 0.03~mag for the \logg=2.25 relation. This
effect would tend to make the slope of the P-L relation slightly steeper,
but the effect is only about half of the estimated uncertainty on
the slopes derived in Sect.\ref{sec.PLrelations}.
We thus confirm the FG97 finding that stars with
very different surface gravities follow the same $F_V,(V-K)$
relation.

We conclude that at the present level of accuracy we
can consider the ISB method to be insensitive to both metallicity and gravity
effects. In the following we will therefore simply apply the same
$F_V - (V-K)_0$ relation to both Galactic and SMC Cepheids and for all periods.

\subsection{The adopted $p$-factor}

\begin{figure}[htp]
\epsfig{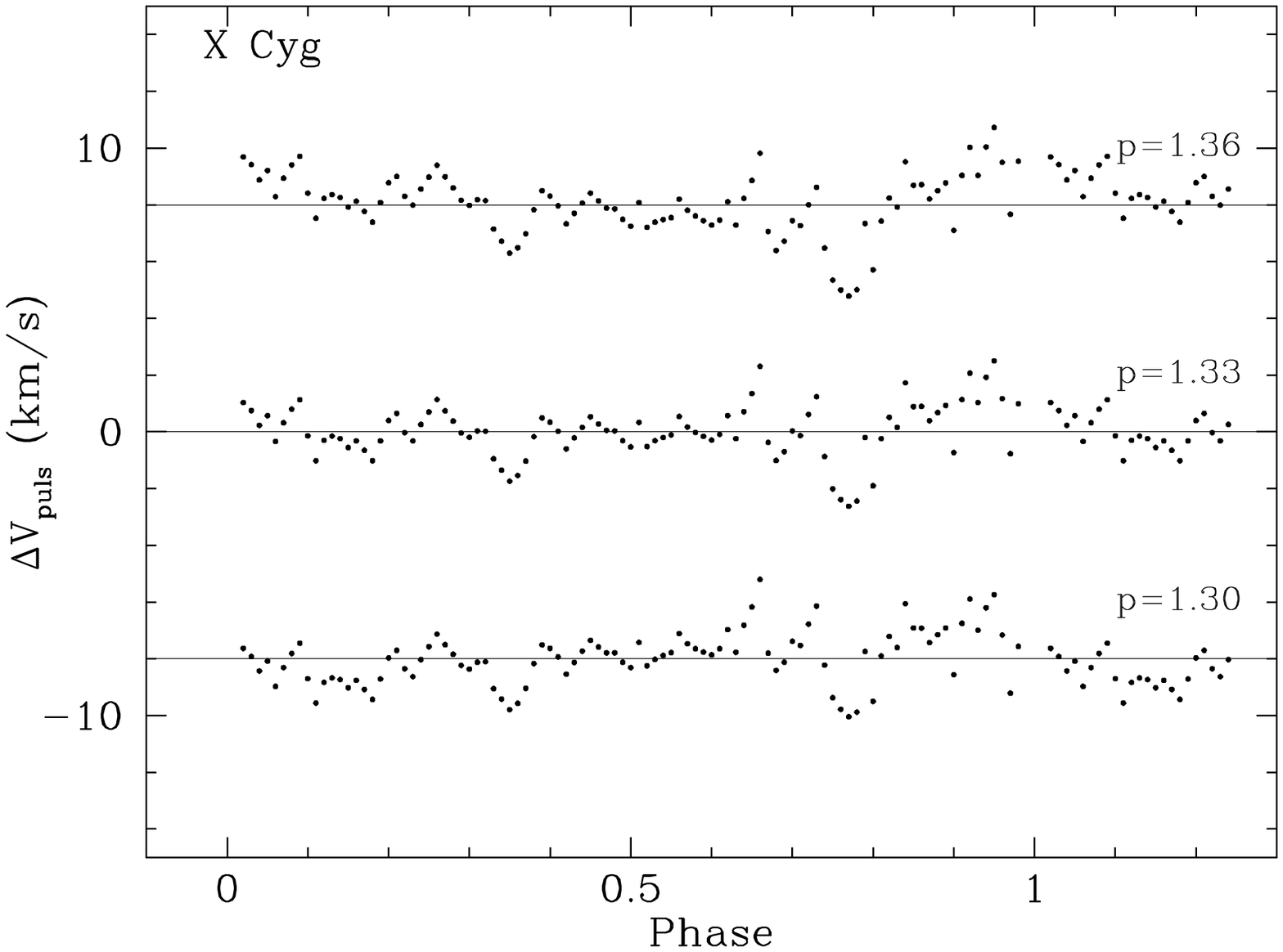}
\figcap{\label{fig.pfactor} The difference in pulsational velocity
between the CfA and CORAVEL data sets for X~Cyg using different
$p$-factors for the CfA data and adopting a $p$-factor of 1.36 for 
the CORAVEL data. The zero-points are arbitrary to separate the three
examples.}
\end{figure}

  The projection factor, $p$, for converting the radial velocities into
pulsational velocities affects the final distances directly as it scales
the radius variation. This means that an overestimate
of the $p$-factor by 1\% leads to an overestimate of the distance by 1\%
(0.02~mag in the distance modulus). Consequently, reliable results can only be
obtained when the $p$-factor is well determined.

  In reality the $p$-factor is not a simple geometric factor but does
also depend on the structure of the stellar atmosphere, and the way the
radial velocity is measured. Parsons (\cite{Parsons72}) computed the
$p$-factor on the basis of line profiles derived from atmosphere models 
to take properly into account the effect of limb darkening, and the
fact that the spectral lines are each formed over a range of optical
depths in the atmosphere. He found values between 1.30 and
1.34 for Cepheids depending on the resolution of the spectrograph.
Hindsley \& Bell (\cite{HindsleyBell86}) investigated the effect of
using a Griffin type photoelectric radial-velocity spectrometer like
CORAVEL (Baranne \cite{Baranne79}) and they found
a somewhat higher value, as well as a weak dependence on stellar surface
temperature. 

  Gieren et al. (\cite{GBM93}) adopted a simple linear weak
dependence on temperature via the period,
\begin{equation}
\label{eq.pfactor}
p = 1.39 - 0.03\log P
\end{equation}
\noindent
based on the discussion of Hindsley \& Bell (\cite{HindsleyBell86},
\cite{HindsleyBell89}), which we will also adopt here.

More recently Sabbey et al. (\cite{Sabbey95}) have performed a careful
observational and theoretical study of the line profiles taking
into account dynamic and non-LTE effects. They find that $p$ is not
entirely constant with phase and that the effect can amount to 7\%
(0.15~mag) in the distances determined with the method. Unfortunately,
additional high resolution and high signal to noise spectra are needed
to determine the phase behaviour of $p$, and for the SMC stars this is
not possible. For lack of data and for consistency between the Galactic
and SMC samples we use a $p$ which is constant for a given star, in
agreement with LTE predictions.  A constant $p$-factor is also supported
by the results of GFG97; for their best-observed stars, like the 6~day
period star U~Sgr, the agreement in shape between the linear displacement
and angular diameter curves is excellent, which should not be the case
if the $p$-factor was significantly phase-dependent.  We note that in
section \ref{sec.results} we find that the match between photometric and
spectroscopic angular diameters breaks down systematically around minimum
radius, exactly where Sabbey et al. (\cite{Sabbey95}) predicts the biggest
deviations from an average $p$ value. In that section we argue that this
phase interval should be disregarded and this will minimize the possible
effect on the derived distances.

A large fraction of the radial-velocity data for the Galactic stars
were obtained using instruments of the photoelectric radial-velocity
spectrometer type for which the previous relation is appropriate. Another
significant fraction, in particular the data for the SMC stars from
Storm et al. (\cite{Storm03}) and the Galactic Cepheids presented
in the Appendix using the CfA spectrographs, are based
on high resolution spectrometers where the radial velocity is determined
by a software cross-correlation using a template in $\log \lambda$ space.

We have the unique opportunity to determine {\em directly} the ratio
between the $p$-factor for CORAVEL and the CfA spectrographs as the stars
X~Cyg and U~Sgr have been observed extensively with both systems.
Both stars have large velocity amplitudes making them excellent
calibration targets.  Each of the data sets are reasonably well
sampled over phase, so we have simply interpolated linearly in each
of the datasets to generate for each star a pair of radial-velocity
curves with a phase spacing of 0.01.  To minimize any possible effect
from amplitude variations for X~Cyg we have only considered the data
obtained in the period between HJD2446000 and HJD2447400.  We have used
the CORAVEL data as the reference and for the CfA velocities we have,
for a number of different $p$ values, determined the differential
pulsation velocity between the two spectrometers, $\Delta
V_{\mbox{\scriptsize rad,p}} =
V_{\mbox{\scriptsize rad,p}}(\mbox{CfA}) 
- V_{\mbox{\scriptsize rad,p}}(\mbox{CORAVEL})$.  For X~Cyg we have plotted
$\Delta V_{\mbox{\scriptsize rad,p}}$ against phase in Fig.\ref{fig.pfactor}.  Assuming that $p$
is constant with phase, $\Delta V_{\mbox{\scriptsize rad,p}}$ should be constant as a function
of phase for the appropriate ratio of $\eta = p_{\mbox{\scriptsize
CfA}} / p_{\mbox{\scriptsize CORAVEL}}$ where $p_{\mbox{\scriptsize
CORAVEL}}$ is computed from Eq.(\ref{eq.pfactor}).  We compute the slope of
$\Delta V_{\mbox{\scriptsize rad,p}}$ versus phase for the phase interval between minimum radial
velocity ($\phi=0.05$) and maximum velocity ($\phi=0.72$) for a range of
different $\eta$ values and determine the $\eta$ value where the slope
is zero to be $\eta=1.34/1.354=0.990\pm0.02$.  The result is robust and
insensitive to the adopted phase interval.  From Fig.\ref{fig.pfactor}
it is obvious that this is a difficult measurement and that relying just
on a few points around maximum and minimum velocity will not lead to a
significant result.  For the short period Cepheid U~Sgr we similarly find
$\eta=1.43/1.365=1.047\pm 0.025$, where the larger error reflects the
fact that the amplitude is smaller and the signal correspondingly
less well constrained.

The difference in $\eta$ for the two stars is hardly significant
and the average value of the two estimates is $\eta=1.019\pm0.02$, which
is very close to unity. It is thus appropriate to use
Eq.(\ref{eq.pfactor}) also for these data.

For the \object{SMC} data set (Storm et al. \cite{Storm03}) the situation
is very similar to that for the CfA data. Again the radial velocities have
been derived from high resolution (0.02~nm) spectra with a high dispersion
(0.2~nm/mm) and a sky exposure was used as the cross-correlation template.
We have rerun the above test for the CfA data using a similar sky
template, and again we found values very close to those for the CORAVEL data.
We thus choose to adopt Eq.(\ref{eq.pfactor}) for the SMC measurements as well.

\subsection{The fitting procedure}



\begin{figure}[htp]
\epsfig{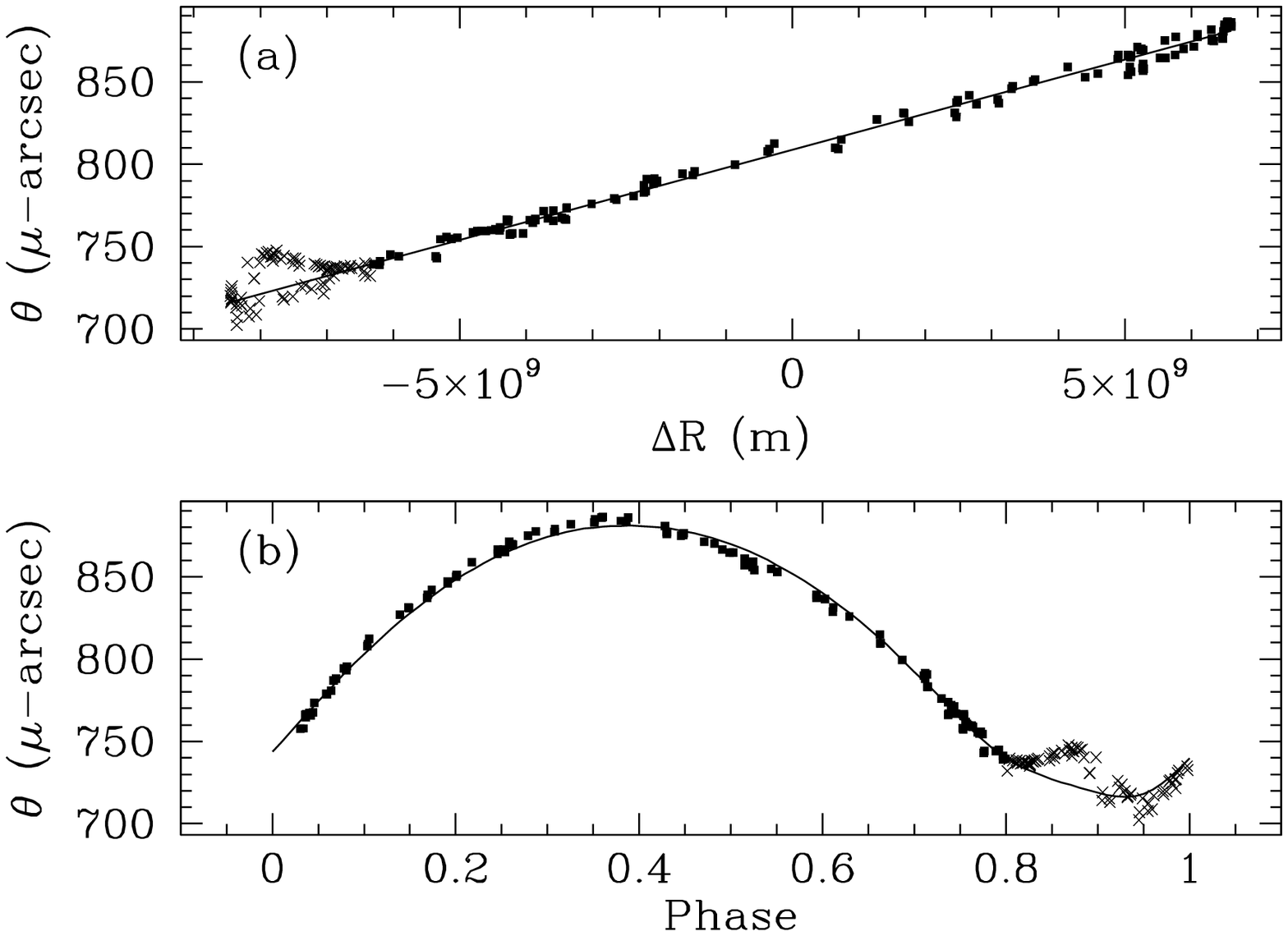}
\figcap{\label{fig.XCyg_fit} 
The points represent the photometrically
determined angular diameters for X~Cyg and the line in panel (a) shows
the bi-sector fit to the filled points.  The curve in panel (b)
delineates the angular diameter obtained from integrating the radial
velocity curve at the derived distance. Crosses represent points which
were eliminated from the fit.}
\end{figure}

There has been extensive discussions in the past (e.g. Laney
\& Stobie \cite{LS95}, GFG97) regarding the proper fitting procedure
for the radius variation versus angular diameter fit. This has been an
issue largely because the scatter in these diagrams, especially for
optical colors, was very significant.
In the near-IR the situation is in general
much better as the scatter is significantly smaller and the choice
of algorithm becomes much less critical.  Since we are interested in
determining the best relation between $\Delta R$ and $\theta$ and not
in predicting $\theta$ from $\Delta R$ or the opposite, we follow the
recommendation by Isobe et al. (\cite{Isobe90}) and use the bi-sector
fit. Such a fit is shown in Fig.\ref{fig.XCyg_fit}a.
Had we chosen to use a linear least squares fit with $\theta$ as the
dependent variable as done by GFG97, we would have obtained distance
moduli which are longer by approximately 0.02~mag.

When we perform the fit we also plot the resulting angular diameter
as a function of phase, based on the resulting distance estimate
(Fig.\ref{fig.XCyg_fit}b). We notice that for many stars the fit in
the phase region between 0.8 and 1.0 is very poor and in almost all
cases the photometric angular diameters show a larger radius than
does the angular diameter curve derived from the radial-velocity
measurements. In addition we see in many cases a significant bump in
this phase interval.  Clearly the method has a problem in this phase
region. Sabbey et al. (\cite{Sabbey95}) and others have argued that the
$p$-factor could be significantly different especially in this phase
region.  Bersier et al. (\cite{Bersier97}) argue that non-LTE effects
and increased microturbulence change the atmosphere structure and this
could make the simple color-surface brightness relation inadequate for
this phase region. As we do not have a simple physical representation
of the situation we prefer to disregard the phase region from 0.8
to 1.0 for all the stars. This does not affect our absolute calibrations
of the Cepheid P-L relations significantly, but the scatter in the P-L
relations becomes slightly smaller.

  For some stars we observe a slight phase shift between the angular
diameter curves derived from photometry and spectroscopy as already
noticed by GFG97. This shift results in a poorly defined linear
relation between $\Delta R$ and $\theta$ which resembles a loop.
The shifts, $\Delta \phi$, have been
determined by minimizing the scatter of the fits and the values are
tabulated in Table \ref{tab.resultsBWIR}. It can be seen that the shifts
are in general small and for most of the  stars it is not even
measurable. We do not have a physical explanation for the effect, but we
do not believe that it is caused by a real phase difference between the
photometric and spectroscopic data. 

The application of the phase shifts marginally reduces the scatter around the
final P-L relation. The effect on the slope of the P-L relation is small,
but of the same size as the internal fitting error. 
It works in the direction of making the slopes in
Tab.\ref{tab.PLrelations} shallower by about 0.15 than what would otherwise
be predicted.
Even though it is unsatisfactory that we do not fully understand the
effect, it is reassuring that the effects on the luminosities are
limited and that only a few stars are affected.

\section{Results}
\label{sec.results}

\begin{table*}
\caption{\label{tab.resultsBWIR}Physical parameters resulting from the
near-IR surface brightness method applied to the samples of Galactic and
SMC Cepheids.}
\scriptsize
\begin{tabular}{r r r r r r r r r r r r r r r}
\hline\hline
\multicolumn{1}{c}{ID} & 
\multicolumn{1}{c}{$\log P$} &
\multicolumn{1}{c}{$(m-M)_0$} &
\multicolumn{1}{c}{$\sigma_{\mbox{\scriptsize (m-M)}}$} &
\multicolumn{1}{c}{$R$} & 
\multicolumn{1}{c}{$\sigma_R$} & 
\multicolumn{1}{c}{$M_B$} &
\multicolumn{1}{c}{$M_V$} &
\multicolumn{1}{c}{$M_I$} &
\multicolumn{1}{c}{$M_J$} &
\multicolumn{1}{c}{$M_H$} &
\multicolumn{1}{c}{$M_K$} &
\multicolumn{1}{c}{$M_W$} &
\multicolumn{1}{c}{$E(B-V)$} &
\multicolumn{1}{c}{$\Delta \phi$} \\
 &  & mag & mag &
\multicolumn{1}{c}{$R_\odot$} & \multicolumn{1}{c}{$R_\odot$} &
\multicolumn{1}{c}{mag} & \multicolumn{1}{c}{mag} &
\multicolumn{1}{c}{mag} & \multicolumn{1}{c}{mag} &
\multicolumn{1}{c}{mag} & \multicolumn{1}{c}{mag} &
\multicolumn{1}{c}{mag} & \multicolumn{1}{c}{mag} & \\
\hline
\multicolumn{15}{c}{Milky Way}\\
\multicolumn{15}{c}{}\\
\object{     SU Cas} &   0.289884 &  8.164 & 0.070 &  29.3 &  0.9 & $-2.725$ & $-3.140$ & $-3.640$ & $-3.925$ & $-4.112$ & $-4.137$ & $-4.391$ &  0.287 & $ 0.000$\\
\object{     EV Sct} &   0.490098 & 11.249 & 0.105 &  34.2 &  1.6 & $-2.878$ & $-3.345$ & $-3.977$ & $-4.230$ & $-4.418$ & $-4.442$ & $-4.923$ &  0.679 & $ 0.045$\\
\object{     BF Oph} &   0.609329 &  9.271 & 0.034 &  32.0 &  0.5 & $-2.134$ & $-2.750$ & $-3.398$ & $-3.844$ & $-4.113$ & $-4.181$ & $-4.372$ &  0.247 & $ 0.035$\\
\object{      T Vel} &   0.666501 &  9.802 & 0.060 &  33.6 &  0.9 & $-2.048$ & $-2.692$ & $-3.370$ & $-3.887$ & $-4.179$ & $-4.260$ & $-4.392$ &  0.281 & $ 0.000$\\
\object{     $\delta$ Cep} &   0.729678 &  7.085 & 0.044 &  42.0 &  0.9 & $-2.870$ & $-3.431$ & $-4.060$ & $-4.471$ & $-4.752$ & $-4.808$ & $-5.007$ &  0.092 & $ 0.000$\\
\object{     CV Mon} &   0.730685 & 10.988 & 0.034 &  40.3 &  0.6 & $-2.460$ & $-3.038$ & $-3.797$ & $-4.263$ & $-4.547$ & $-4.647$ & $-4.932$ &  0.714 & $ 0.015$\\
\object{      V Cen} &   0.739882 &  9.175 & 0.063 &  42.0 &  1.2 & $-2.714$ & $-3.295$ & $-3.957$ & $-4.414$ & $-4.694$ & $-4.770$ & $-4.950$ &  0.289 & $ 0.000$\\
\object{     BB Sgr} &   0.821971 &  9.519 & 0.028 &  49.8 &  0.6 & $-2.817$ & $-3.518$ & $-4.262$ & $-4.724$ & $-5.029$ & $-5.102$ & $-5.382$ &  0.284 & $-0.035$\\
\object{      U Sgr} &   0.828997 &  8.837 & 0.021 &  47.7 &  0.5 & $-2.785$ & $-3.477$ & $-4.213$ & $-4.668$ & $-4.950$ & $-5.023$ & $-5.317$ &  0.403 & $ 0.000$\\
\object{    $\eta$ Aql} &   0.855930 &  6.990 & 0.045 &  48.4 &  1.0 & $-2.946$ & $-3.581$ & $-4.271$ & $-4.717$ & $-5.012$ & $-5.073$ & $-5.310$ &  0.149 & $ 0.000$\\
\multicolumn{15}{c}{}\\
\object{      S Nor} &   0.989194 &  9.908 & 0.032 &  70.7 &  1.0 & $-3.345$ & $-4.101$ & $-4.856$ & $-5.411$ & $-5.734$ & $-5.818$ & $-5.995$ &  0.189 & $ 0.000$\\
\object{     XX Cen} &   1.039548 & 11.114 & 0.022 &  69.4 &  0.7 & $-3.430$ & $-4.154$ & $-4.897$ & $-5.415$ & $-5.719$ & $-5.802$ & $-6.015$ &  0.260 & $-0.040$\\
\object{   V340 Nor} &   1.052579 & 11.145 & 0.185 &  67.1 &  5.7 & $-2.985$ & $-3.814$ & $-4.679$ & $-5.225$ & $-5.577$ & $-5.670$ & $-5.980$ &  0.315 & $ 0.000$\\
\object{     UU Mus} &   1.065819 & 12.590 & 0.084 &  74.1 &  2.9 & $-3.426$ & $-4.159$ & $-4.926$ & $-5.495$ & $-5.813$ & $-5.905$ & $-6.079$ &  0.413 & $-0.005$\\
\object{      U Nor} &   1.101875 & 10.716 & 0.060 &  76.3 &  2.1 & $-3.713$ & $-4.415$ & $-5.141$ & $-5.645$ & $-5.928$ & $-6.017$ & $-6.225$ &  0.892 & $ 0.000$\\
\object{     BN Pup} &   1.135867 & 12.951 & 0.050 &  83.3 &  1.9 & $-3.765$ & $-4.513$ & $-5.270$ & $-5.780$ & $-6.100$ & $-6.182$ & $-6.408$ &  0.438 & $ 0.000$\\
\object{     LS Pup} &   1.150646 & 13.554 & 0.056 &  90.1 &  2.3 & $-3.926$ & $-4.685$ & $-5.433$ & $-5.953$ & $-6.281$ & $-6.357$ & $-6.554$ &  0.478 & $ 0.000$\\
\object{     VW Cen} &   1.177138 & 12.804 & 0.039 &  86.7 &  1.6 & $-3.147$ & $-4.037$ & $-4.932$ & $-5.630$ & $-6.022$ & $-6.136$ & $-6.275$ &  0.448 & $ 0.000$\\
\object{      X Cyg} &   1.214482 & 10.420 & 0.018 & 105.6 &  0.9 & $-4.123$ & $-4.991$ & $-5.768$ & $-6.273$ & $-6.615$ & $-6.691$ & $-6.939$ &  0.288 & $ 0.000$\\
\object{     VY Car} &   1.276818 & 11.499 & 0.022 & 112.8 &  1.1 & $-3.929$ & $-4.846$ & $-5.702$ & $-6.324$ & $-6.679$ & $-6.783$ & $-6.994$ &  0.243 & $-0.020$\\
\multicolumn{15}{c}{}\\
\object{     RY Sco} &   1.307927 & 10.514 & 0.034 &  99.9 &  1.6 & $-4.394$ & $-5.060$ & $-5.806$ & $-6.269$ & $-6.540$ & $-6.623$ & $-6.923$ &  0.777 & $ 0.000$\\
\object{     RZ Vel} &   1.309564 & 11.021 & 0.029 & 114.8 &  1.5 & $-4.250$ & $-5.042$ & $-5.826$ & $-6.408$ & $-6.735$ & $-6.827$ & $-7.000$ &  0.335 & $ 0.000$\\
\object{     WZ Sgr} &   1.339443 & 11.287 & 0.047 & 121.8 &  2.6 & $-3.874$ & $-4.801$ & $-5.721$ & $-6.381$ & $-6.765$ & $-6.881$ & $-7.103$ &  0.467 & $ 0.000$\\
\object{     WZ Car} &   1.361977 & 12.918 & 0.066 & 112.1 &  3.4 & $-4.142$ & $-4.918$ & $-5.718$ & $-6.322$ & $-6.661$ & $-6.745$ & $-6.922$ &  0.384 & $ 0.000$\\
\object{     VZ Pup} &   1.364945 & 13.080 & 0.056 &  96.9 &  2.5 & $-4.321$ & $-5.009$ & $-5.721$ & $-6.189$ & $-6.491$ & $-6.553$ & $-6.781$ &  0.471 & $ 0.000$\\
\object{     SW Vel} &   1.370016 & 11.995 & 0.025 & 117.3 &  1.4 & $-4.211$ & $-5.019$ & $-5.844$ & $-6.444$ & $-6.786$ & $-6.890$ & $-7.084$ &  0.349 & $-0.020$\\
\object{      T Mon} &   1.431915 & 10.815 & 0.055 & 149.5 &  3.8 & $-4.403$ & $-5.372$ & $-6.247$ & $-6.892$ & $-7.274$ & $-7.375$ & $-7.564$ &  0.209 & $ 0.000$\\
\object{     RY Vel} &   1.449158 & 12.019 & 0.032 & 139.9 &  2.1 & $-4.693$ & $-5.501$ & $-6.302$ & $-6.885$ & $-7.183$ & $-7.277$ & $-7.506$ &  0.562 & $-0.005$\\
\object{     AQ Pup} &   1.478624 & 12.522 & 0.045 & 147.9 &  3.1 & $-4.649$ & $-5.513$ & $-6.407$ & $-6.949$ & $-7.301$ & $-7.403$ & $-7.745$ &  0.512 & $-0.055$\\
\object{     KN Cen} &   1.531857 & 13.124 & 0.045 & 185.8 &  3.9 & $-5.642$ & $-6.328$ & $-6.975$ & $-7.506$ & $-7.836$ & $-7.936$ & $-7.937$ &  0.926 & $ 0.005$\\
\multicolumn{15}{c}{}\\
\object{      l Car} &   1.550855 &  8.990 & 0.032 & 201.8 &  2.9 & $-4.712$ & $-5.821$ & $-6.772$ & $-7.454$ & $-7.871$ & $-7.965$ & $-8.205$ &  0.170 & $-0.040$\\
\object{      U Car} &   1.589083 & 10.973 & 0.032 & 161.6 &  2.4 & $-4.725$ & $-5.617$ & $-6.480$ & $-7.104$ & $-7.452$ & $-7.559$ & $-7.779$ &  0.283 & $-0.050$\\
\object{     RS Pup} &   1.617420 & 11.561 & 0.064 & 208.0 &  6.1 & $-5.047$ & $-6.015$ & $-6.962$ & $-7.599$ & $-7.967$ & $-8.084$ & $-8.384$ &  0.446 & $ 0.000$\\
\object{     SV Vul} &   1.653162 & 12.102 & 0.037 & 224.0 &  3.8 & $-5.876$ & $-6.752$ & $-7.567$ & $-8.001$ & $-8.314$ & $-8.373$ & $-8.785$ &  0.570 & $-0.045$\\
\multicolumn{15}{c}{SMC}\\
\multicolumn{15}{c}{}\\
\object{    HV 1345} &   1.129638 & 18.828 & 0.081 &  78.4 &  2.9 & $-3.502$ & $-4.166$ & $-4.926$ & $-5.515$ & $     $ & $-5.998$ & $-6.072$ &  0.030 & $ 0.000$\\
\object{    HV 1335} &   1.157807 & 18.875 & 0.082 &  72.6 &  2.8 & $-3.808$ & $-4.365$ & $-5.045$ & $-5.545$ & $     $ & $-5.933$ & $-6.070$ &  0.090 & $ 0.000$\\
\object{    HV 1328} &   1.199645 & 18.732 & 0.087 &  80.4 &  3.2 & $-4.051$ & $-4.617$ & $-5.295$ & $-5.747$ & $     $ & $-6.164$ & $-6.319$ &  0.000 & $ 0.000$\\
\object{    HV 1333} &   1.212014 & 19.388 & 0.080 & 100.3 &  3.7 & $-4.271$ & $-4.913$ & $-5.660$ & $-6.171$ & $     $ & $-6.601$ & $-6.787$ &  0.070 & $ 0.000$\\
\object{     HV 822} &   1.223810 & 19.091 & 0.081 &  97.0 &  3.6 & $-3.934$ & $-4.673$ & $-5.495$ & $-6.014$ & $     $ & $-6.473$ & $-6.729$ &  0.030 & $ 0.000$\\
\hline
\end{tabular}
\end{table*}

  We have applied the procedure described in the previous section to
both the sample of Galactic Cepheids and SMC Cepheids. The resulting radii,
absolute magnitudes, and distances have been tabulated in 
Table \ref{tab.resultsBWIR} together with the adopted reddenings.
The quoted errors in the table are 1 $\sigma$ statistical errors.

\subsection{The luminosity zero-point}

  For the Cepheids which are considered members of open clusters or
associations we can compare the Cepheid distances which we have derived
using the infrared surface brightness method with the distances found by
ZAMS fitting. This is equivalent to comparing the luminosity zero-points
of the two methods.
Turner \& Burke (\cite{TB02}) have published the latest and most
extensive list of ZAMS based distances to cluster and association
Cepheids including a total of 46 stars.

Eighteen of these stars are also present in our sample
and can be used to compare the two methods directly. U~Car and SV~Vul
are outside the period interval which we have adopted for computing our
P-L relations, so they have been eliminated as has AQ~Pup which deviates by
$0.79\pm0.11$~mag and which is considered an uncertain cluster member.
Based on the remaining fifteen stars we find a weighted mean distance
difference of
\begin{equation}
(m-M)_{\mbox{\scriptsize ISB}}-(m-M)_{\mbox{\scriptsize ZAMS}}=+0.01\pm0.06
\end{equation}
\noindent
with a dispersion of $\sigma =0.23$. If we remove three additional
outliers ($\delta$~Cep, $0.37\pm0.11$; BB~Sgr, $0.49\pm0.08$; and U~Car,
$-0.45\pm0.05$) for which the case for cluster membership is also not
considered very strong the dispersion decreases to $\sigma=0.12$ and the
mean value changes marginally to $+0.05\pm0.04$.

Feast (\cite{Feast99}) points out that the
Hipparcos results indicate that the ZAMS fitting method is probably
less well understood than previously assumed and that the zero-point can
be uncertain by perhaps 0.2~mag. On the other hand we can conclude that
the two methods are in excellent agreement and we take this as further
support for the correctness of the luminosity zero-point for the ISB method
and thereby for the zero-point of the resulting P-L relations.

 Fouqu\'e et al. (\cite{FSG03}) have compared the latest
Hipparcos determination of the zero-point of the P-L relations with the
results from the surface brightness analysis of a sample of Galactic 
Cepheids. They find generally good agreement, lending additional
support to the calibration of Fouqu\'e \& Gieren (\cite{FG97}) and the
zero-point of the resulting P-L relations.

\subsection{The P-L relations}
\label{sec.PLrelations}

\begin{figure}[htp]
\epsfig{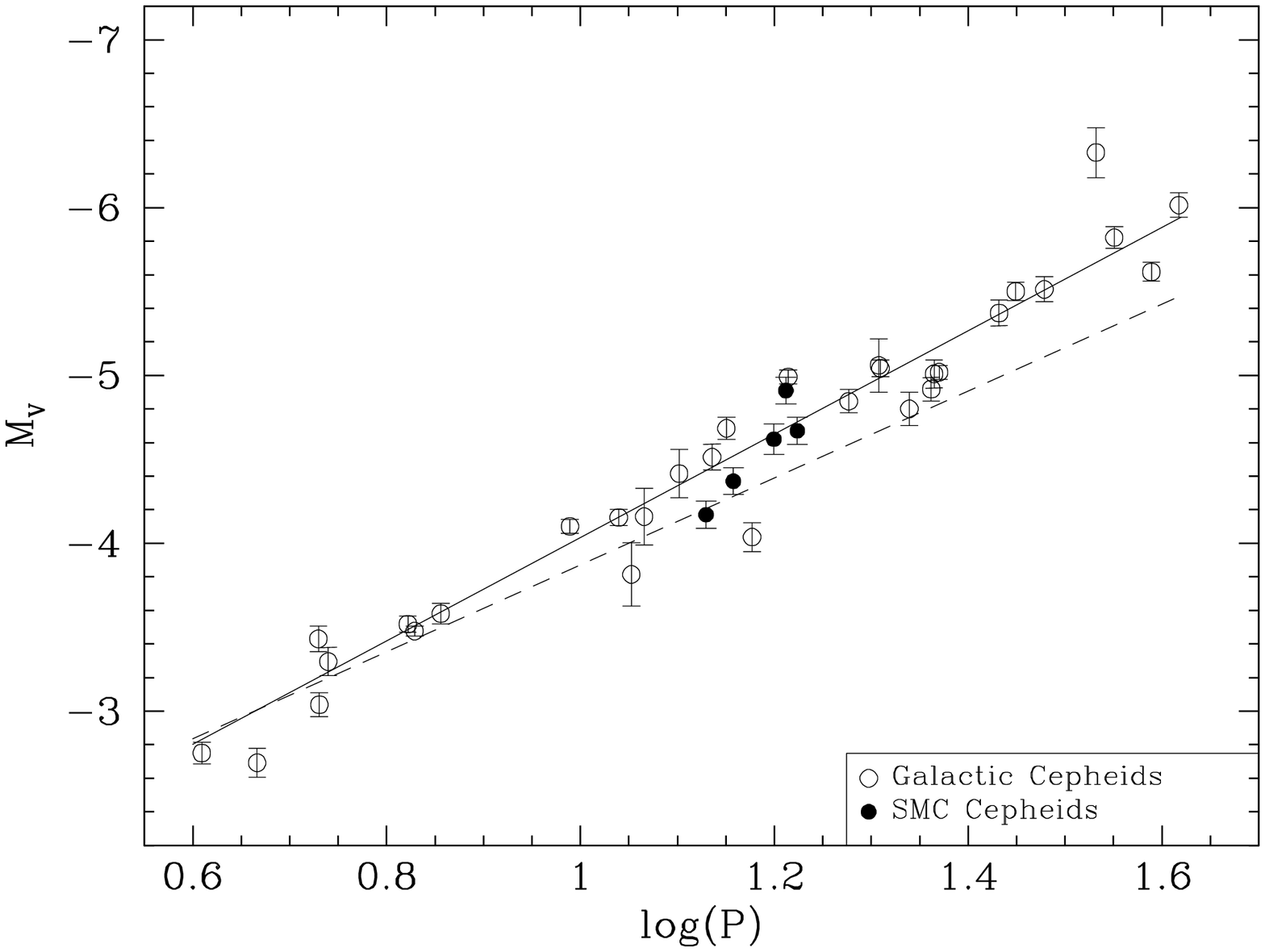}
\figcap{\label{fig.logPMv} The absolute $V$ magnitudes for the Galactic
and SMC Cepheids as derived from the Baade-Wesselink analysis plotted
against $\log P$.  The dashed
line represents the SMC relation based on OGLE data shifted to the 
adopted SMC distance (see Eq.(\ref{eq.logPMv_SMC})).}
\end{figure}

\begin{figure}[htp]
\epsfig{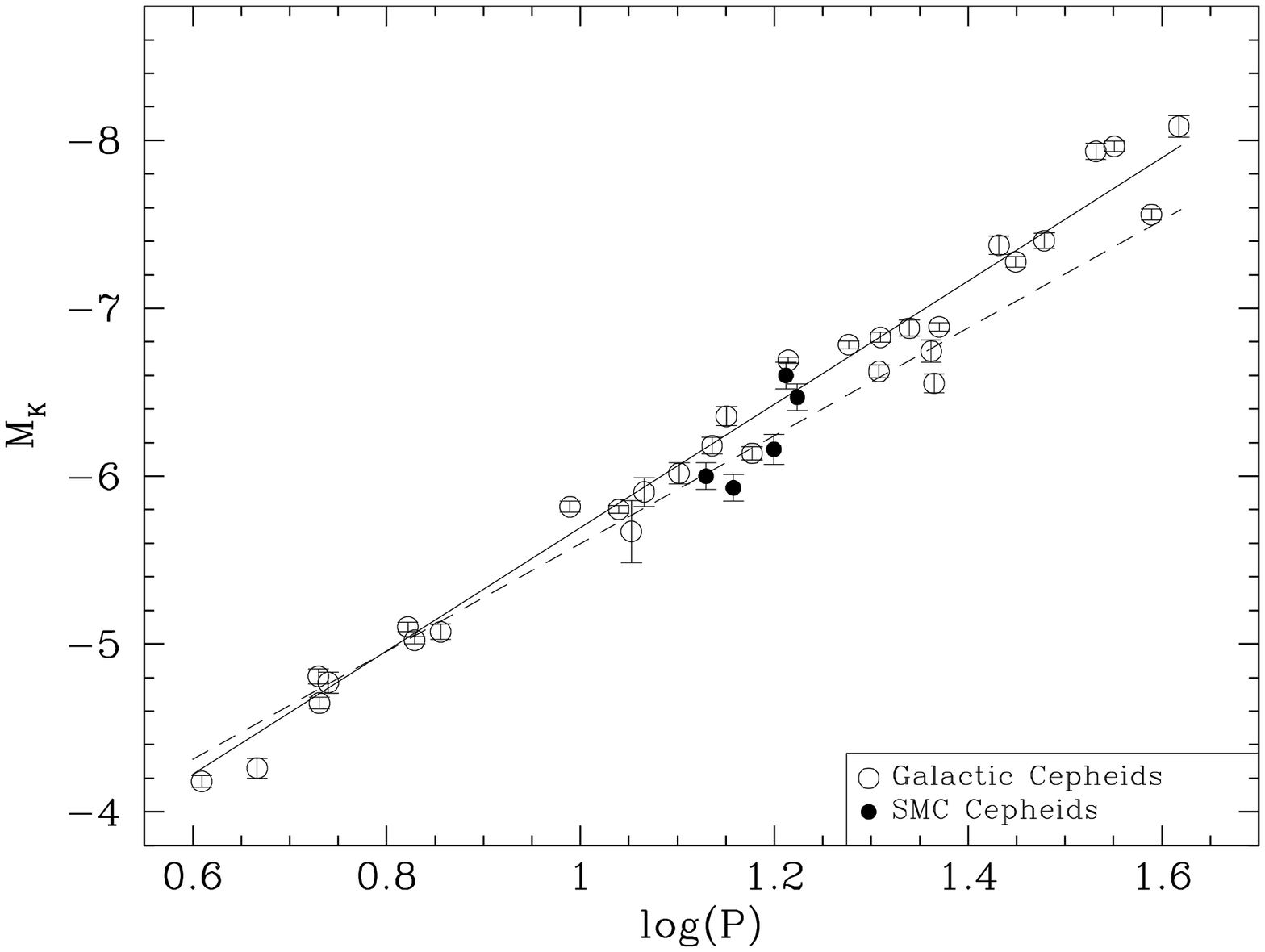}
\figcap{\label{fig.logPMk} The absolute $K$ magnitudes for the Galactic
and SMC Cepheids as derived from the Baade-Wesselink analysis plotted
against $\log P$. The dashed line represents the SMC relation based on
Eq.(\ref{eq.PLkSMC}) and shifted to the adopted SMC distance.}
\end{figure}

\begin{figure}[htp]
\epsfig{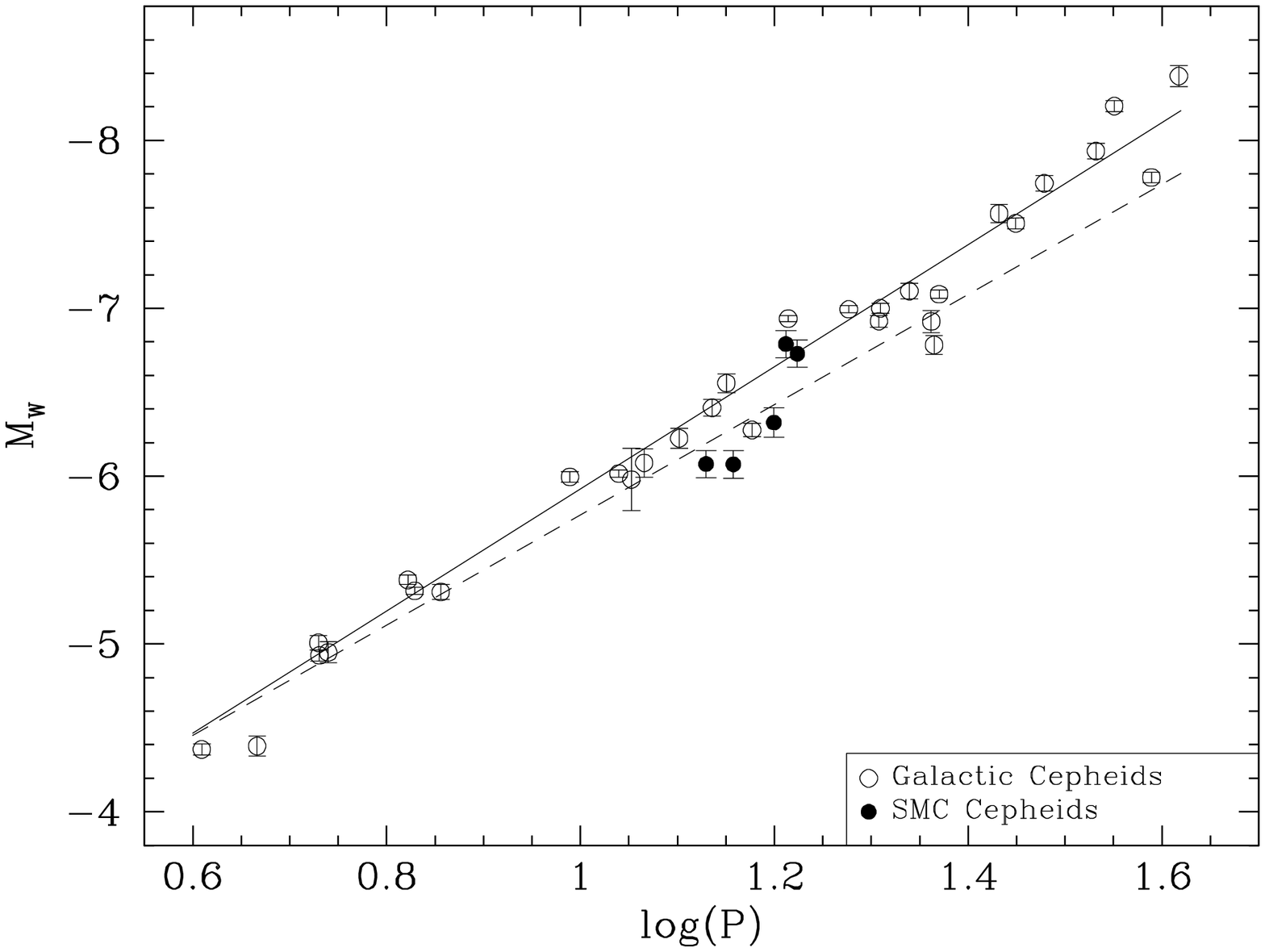}
\figcap{\label{fig.logPW} The Wesenheit magnitudes for the Galactic
and SMC Cepheids as derived from the Baade-Wesselink analysis plotted
against $\log P$. The best linear fit has been overplotted. 
The dashed
line represents the SMC relation based on OGLE data shifted to the 
adopted SMC distance (see Eq.(\ref{eq.logPMw_SMC})).}
\end{figure}

  Based on the absolute magnitudes for the 32 Galactic
Cepheids in the relevant period interval ($0.60 < \log P < 1.62$)
we re-derive the P-L relation in these bands in the form:
\begin{equation}
M_x = \alpha (\log P - 1.00) + \beta
\end{equation}
\noindent
and 
\begin{equation}
M_x = \alpha (\log P - 1.18) + \gamma
\end{equation}
\noindent
where $M_x$ is the absolute magnitude in band $x$ and the offset in
$\log P$ of 1.18 corresponds to the mean period of the SMC sample. 
The fit of the latter form provides a formal estimate of the uncertainty in the
magnitude at the average period of our SMC sample.  The fitted coefficients
and their error estimates are tabulated in Table \ref{tab.PLrelations}.

\begin{table}
\caption{\label{tab.PLrelations} The Period-Luminosity relations for the
Galactic sample in different bands and the RMS around the fit.}
\begin{tabular}{r l l l r}
\hline\hline
Band & \multicolumn{1}{c}{$\alpha$}  & \multicolumn{1}{c}{$\beta$} &
\multicolumn{1}{c}{$\gamma$} & \multicolumn{1}{c}{RMS}\\
\hline
$B$ & $-2.74\pm0.12$ & $-3.30\pm0.03$ & $-3.79\pm0.03$ & 0.18\\
$V$ & $-3.08\pm0.11$ & $-4.03\pm0.04$ & $-4.59\pm0.03$ & 0.17\\
$I$ & $-3.30\pm0.10$ & $-4.79\pm0.03$ & $-5.38\pm0.03$ & 0.16\\
$J$ & $-3.53\pm0.10$ & $-5.30\pm0.03$ & $-5.93\pm0.03$ & 0.15\\
$H$ & $-3.63\pm0.11$ & $-5.61\pm0.03$ & $-6.26\pm0.03$ & 0.16\\
$K$ & $-3.67\pm0.12$ & $-5.69\pm0.03$ & $-6.35\pm0.03$ & 0.15\\
$W$ & $-3.63\pm0.11$ & $-5.92\pm0.04$ & $-6.58\pm0.03$ & 0.16\\
\hline
\end{tabular}
\end{table}

We have also determined the P-L relation for the Wesenheit function, $W$,
as used by the HST Key Project (Freedman et al. \cite{Freedman01}) and
adopting our value of $R_W=2.51$ from Sect.\ref{sec.GalRed}.
We have then defined $M_W = W_{VI} - (m-M)_0$ for our Galactic stars, where
the distance modulus comes from our reddening corrected surface brightness
analysis in Table \ref{tab.resultsBWIR}, and proceeded as for the other
bands. 

The resulting relations provide a direct starting point for determining
distances to the HST Key Project galaxies without any assumptions
regarding the LMC distance and, as the metallicity difference
between the Galaxy and the Key Project galaxies is, on average, smaller
than between the LMC and the target galaxies, any metallicity effect will
have a smaller effect on the derived distances. The relation based on
the Wesenheit function allows us to derive $V$ and $I$ based distances
to the Key Project galaxies using the Wesenheit function to correct for
reddening in the external galaxies. This approach requires a
zero-point for our reddening scale for the Galactic Cepheids, whereas
the Key Project (Freedman et al. \cite{Freedman01}) can work purely
differentially. We, however, have the advantage of being able to determine
distances without assuming a distance to the LMC and with a smaller
metallicity difference between the galaxies and the reference (the
Galaxy).

  We note that all the P-L relations which we have derived here are, as
expected, in very good agreement with the relations derived by 
Gieren et al. (\cite{GFG98})
which used the same method but a slightly smaller data set.

  The absolute magnitudes versus $\log P$ and the fitted relations are
shown in Figs.\ref{fig.logPMv}-\ref{fig.logPW}.

\subsection{The Period-Radius relation}
\label{sec.logPR}

\begin{figure}[htp]
\epsfig{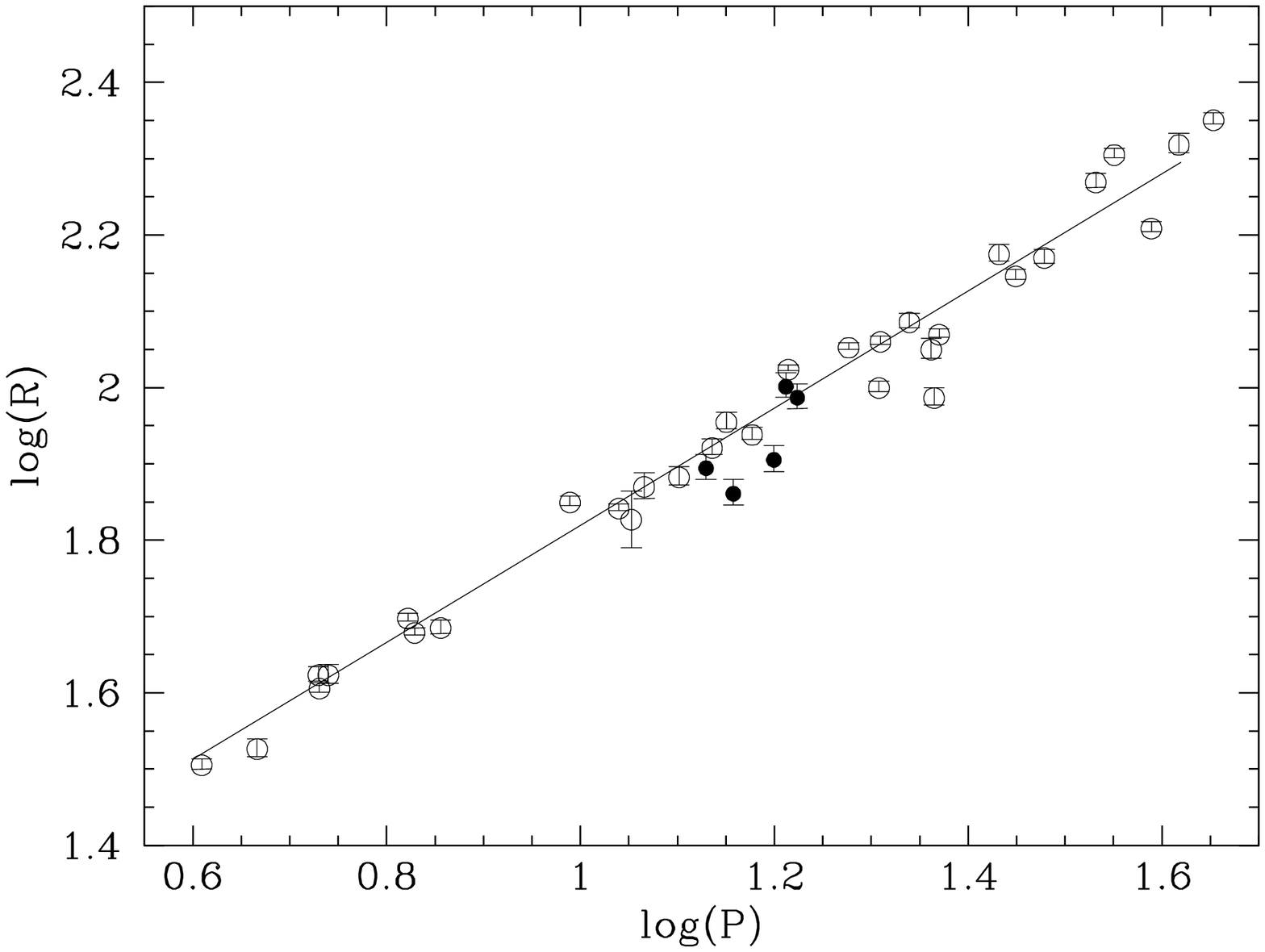}
\figcap{\label{fig.logPR} The radii for the Galactic and SMC Cepheids as
derived from the infrared surface brightness method plotted against $\log P$.
The best fit straight line to the Galactic stars has been overplotted.}
\end{figure}

In Fig.\ref{fig.logPR} we have plotted the period-radius diagram from our
data.  The Galactic Cepheids define a linear relation
\begin{equation}
\log R = 0.77(\pm0.02) \log P + 1.05(\pm0.03)
\end{equation}
\noindent
which has been overplotted on the data points in the figure.
This relation is, as expected, in excellent agreement with that of 
Gieren et al.(\cite{GFG98}).

  It can be seen that the \object{SMC} stars in our sample are on average
slightly smaller than the corresponding Galactic stars. Three of the
stars are very close to the Galactic locus and \object{HV1335} exhibits
the largest deviation followed by \object{HV1328}. \object{HV1328}
however, is already expected to be an extreme case as it is very close
to the blue edge of the instability strip and the radial-velocity
amplitude is extremely low ($A_{\mbox{\scriptsize RV}}=23$~km/s, Storm
et al. \cite{Storm03}) for its period.

\subsection{The distance to the SMC}
\label{Sect.SMCdist}

\begin{figure}[htp]
\epsfig{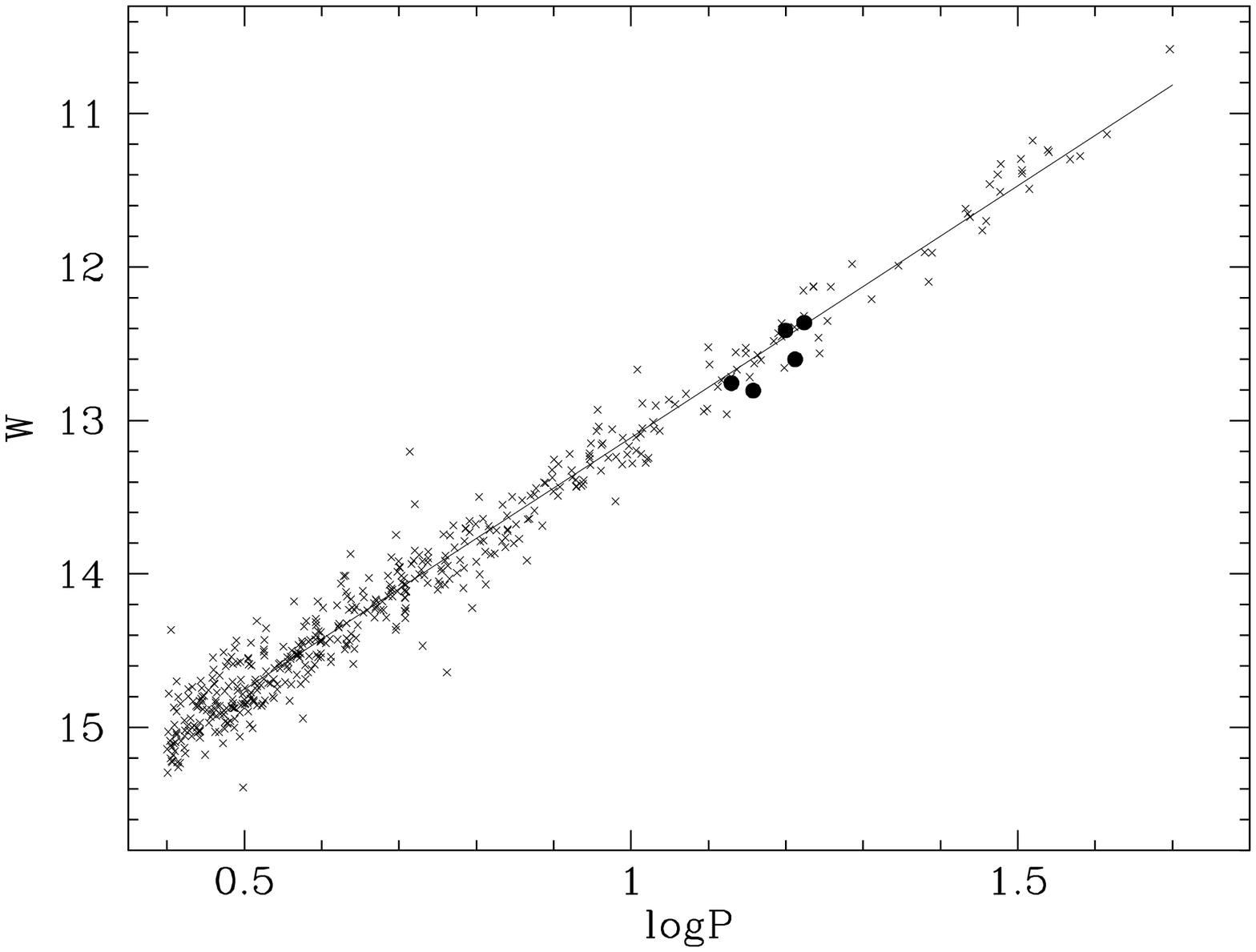}
\figcap{\label{fig.OGLE_PLW} The $W$-indices for the OGLE-2
sample plotted agains $\log P$. Our sample of stars have been overplotted
as filled circles. The straight line shows the linear regression to the
OGLE-2 data.}
\end{figure}

  The individual distance moduli to the present sample of SMC stars, 
as measured by the surface brightness method, are tabulated
in Table \ref{tab.resultsBWIR}, and the weighted mean distance modulus is
$(m-M)_0=19.00\pm 0.12$.
This distance estimate is very robust to errors in the reddening
estimate. Assuming a mean reddening of $0.00$ or $0.12$ would lead to
moduli of $18.96$ and $19.03$ respectively.

  To obtain the distance to the SMC itself, we formally have to take into
consideration the depth effects in the SMC. Laney \& Stobie
(\cite{LS94}) used the simple geometrical model of Caldwell \& Laney
(\cite{CL91}) to derive corrections to the moduli of SMC Cepheids,
including four of the stars in our sample. They found the values
$-0.142$ for HV1328, $-0.126$ for HV1333, $-0.122$ for HV1335, and
$-0.098$ for HV1345. We have used the same method
to derive a correction for the last star, \object{HV~822}, for which we
find a value of $-0.093$. We have also looked at the OGLE-2 $W-\log P$
relation and found an effect of similar size and direction as that
determined by Caldwell \& Laney (\cite{CL91}). However, as the OGLE sample
does not extend as far from the SMC center as the stars used by Caldwell
and Laney we adopt the latter values as they are based on a wider
spatial baseline. The mean offset for our five stars is $-0.116\pm 0.009$~mag
in the sense that the distance to the bulk of the SMC is shorter by this
amount than the distance to these five stars.  In Fig.\ref{fig.OGLE_PLW}
the $W$-$\log P$ relation for the OGLE-2 sample is shown with our sample
overplotted.
The $W$-index for the OGLE sample has been computed using
our value of $R_W=2.51$ for consistency.
It is evident that our sample is slightly fainter than the
ridge line of the OGLE-2 sample, by approximately 0.1~mag, further
justifying the application of the depth correction. 

Applying the individual offsets to the moduli in Table \ref{tab.resultsBWIR}
and computing the weighted mean value gives an SMC distance modulus of $(m-M)_0
= 18.88\pm0.12$~mag.  Due to the limited sample size and the fact that
the depth effect is a statistical effect of a size comparable to the
spread around the ridge line, we add in quadrature an additional error
of $0.07$mag to the resulting mean value leading to a best estimate of the
SMC distance modulus of $(m-M)_0 = 18.88\pm0.14$~mag.

\subsection{The slopes of the P-L relations}

We can compare the slopes for the Galactic P-L relations which we have
derived in Sect.\ref{sec.PLrelations} with the slopes derived for the SMC
Cepheids on the basis of the dereddened data on fundamental mode Cepheids
in the OGLE-2 data base (Udalski et al.  \cite{Udalski99b}). Fouqu\'e et
al. (\cite{FSG03}) found that the OGLE-2 reddenings for LMC Cepheids
were too large when compared to other calibrations.  We have included
the OGLE-2 reddening estimates for the three stars in common with our
sample in Table \ref{tab.BVIred}. The values are slightly larger than
the Laney \& Stobie (\cite{LS94}) values and also larger than our $BVI$
based estimates which we have adopted for
our sample. To formally bring the two samples on the same reddening system we
have simply offset the reddenings for the OGLE-2 stars by $-0.025$~mag.
We then extract the stars in the period interval $0.4 < \log P < 1.7$
and fit a P-L relation using linear regression after rejecting between
one and five outlying points for each relation. The resulting fits are
tabulated in Table \ref{tab.SMCPLrelations}. In the optical bands the
scatter is rather large but in the $W$-index, which is insensitive to
reddening, the RMS around the fit decreases to 0.14~mag, even smaller
than for our Galactic sample. This suggests that the increased width
of the other P-L relations in the optical bands are not introduced by
depth effects but rather by uncertainties in the individual reddening
estimates.

In the $K$~band we can use the $K$~band P-L
relation from Groenewegen (\cite{Groen00}):
\begin{equation}
\label{eq.PLkSMC}
K_0 = -3.212\pm0.033 \log P +16.494\pm0.026
\end{equation}
\noindent
This relation is based on
single or few phase measurements from the DENIS (Epchtein et al.
\cite{Epchtein99}) and 2MASS (Jarrett et al. \cite{Jarrett00}) surveys.  

\begin{table}
\caption{\label{tab.SMCPLrelations} The Period-Luminosity relations for the
488 OGLE-2 SMC Cepheids in the period interval $0.4 < \log P < 1.7$ in
different bands. A few outliers have been rejected before the fits.
A reddening of $E(B-V)=E(B-V)_{\mbox{\scriptsize OGLE}}-0.025$ has been
assumed.}
\begin{tabular}{c l l r}
\hline\hline
Band & \multicolumn{1}{c}{$\alpha$}  & \multicolumn{1}{c}{$\beta$} & \multicolumn{1}{c}{RMS}\\
\hline
$V$  & $-2.590\pm0.047$ & $+15.010\pm0.019$ & 0.288 \\
$I$  & $-2.865\pm0.036$ & $+14.252\pm0.014$ & 0.220 \\
$W$  & $-3.283\pm0.023$ & $+13.112\pm0.009$ & 0.144 \\
\hline
\end{tabular}
\end{table}

\begin{table}
\caption{\label{tab.SMCPLMWrelations} The Period-Luminosity relations for the
Galactic Cepheids but forcing the slope to be the one 
derived from the SMC sample.}
\begin{tabular}{c l l r}
\hline\hline
Band & \multicolumn{1}{c}{$\alpha$}  & \multicolumn{1}{c}{$\beta$} & \multicolumn{1}{c}{RMS}\\
\hline
$V$  & $-2.590$ & $-4.103\pm0.046$ & 0.22 \\
$I$  & $-2.865$ & $-4.854\pm0.043$ & 0.21 \\
$W$  & $-3.283$ & $-5.990\pm0.042$ & 0.20 \\
$K$  & $-3.212$ & $-5.775\pm0.044$ & 0.20 \\
\hline
\end{tabular}
\end{table}

Adopting our best estimate of the SMC distance of $(m-M)_0=18.88$ and
combining this with the OGLE-2 relations from
Table \ref{tab.SMCPLrelations}, and the $K$~band relation from Groenewegen
(\cite{Groen00}) we derive the relations
\begin{eqnarray}
\label{eq.logPMv_SMC}
M_V\mbox{(SMC)} & = & -2.590(\logP-1)-3.87\\
\label{eq.logPMi_SMC}
M_I\mbox{(SMC)} & = & -2.865(\logP-1)-4.63\\
\label{eq.logPMw_SMC}
M_W\mbox{(SMC)} & = & -3.283(\logP-1)-5.77\\
\label{eq.logPMk_SMC}
M_K\mbox{(SMC)} & = & -3.212(\logP-1)-5.60
\end{eqnarray}

The dashed lines in Figs.\ref{fig.logPMv}-\ref{fig.logPW} show these
relations on top of the Galactic relations.  It is obvious that the
slopes of the SMC relations are less steep than the relations based on the
Galactic sample.  The difference is significant for all the photometric
bands, and could be an effect of the difference in metallicity between
the two samples. Evidence for such a significant
dependence of the P-L slopes on metallicity has also recently been found by
Tammann et al. (\cite{Tammann03}).

  If the difference in slope is real and indeed dependent on metallicity,
then it will have very significant consequences for using the P-L relation as a 
distance estimator. In the relevant period range from $\log P=0.6$ to
$\log P=1.6$ the effect in the $W$-index would amount to more than
0.3~mag as the slopes differ by $-3.283-(-3.63)=0.35$. 

Udalski
et al. (\cite{Udalski01}) find that the slopes are similar for the three
low metallicity galaxies IC1613 ($\FeH=-1.0$), SMC ($\FeH=-0.7$), and LMC
($\FeH=-0.3$) and they argue that there is no metallicity effect on the
P-L relations. However, they do not have a solar-metallicity galaxy in
their sample so if we are really seeing a metallicity effect, then it
might not be a simple linear effect in $\FeH$.

  It should be kept in mind that for periods shorter than about 10 days
there is a significant risk for contaminating the sample of fundamental
mode pulsators with first overtone pulsators which would be brighter at
a given period. Udalski et al. (\cite{Udalski99b}) have used Fourier
parameters to discriminate between pulsation modes which works well for
most periods, except in the interval $\logP =0.6$ to $0.8$.
This could, in principle, change the observed slope. For
the use for extra-galactic distances short-period Cepheids are currently
not so interesting as they are too faint, so we can conveniently
circumvent this possible problem by simply disregarding stars with
periods shorter than 10 days ($\log P = 1.$). Similarly, we make a cut
at the long period end at $\log P =1.5$ to avoid possible problems with
the most luminous stars which are not relevant for our 15~day period
stars.
The slopes for the SMC and Galactic samples indeed become more similar,
but the error estimates also become much larger due to the more limited
baseline, so the significance of any difference in slopes becomes very
small. The luminosity differences which we will derive in the next section
changes only by a few hundredth of magnitude, so we have decided to simply
employ the full relations from above.

\subsection{The metallicity effect on Cepheid luminosities}

We can now investigate the metallicity effect on the
zero-point of the P-L relation using three different approaches.

In the first approach we adopt the slopes of the SMC relations as the
``true'' slopes and fit only the zero-point for the Galactic sample. The
resulting zero-points are tabulated in Table \ref{tab.SMCPLMWrelations}.
For the optical bands we
can compare these zero-points with the zero-points based on the OGLE-2
sample for an assumed distance modulus of $(m-M)_0(\mbox{SMC})=18.88$
as given in the equations Eq.(\ref{eq.logPMv_SMC}-\ref{eq.logPMw_SMC}).  The
difference $\Delta M = M_{\mbox{\scriptsize MW}} - M_{\mbox{\scriptsize
SMC}}$ is tabulated in column 2 of Table \ref{tab.dM}. The effect is about
$-0.22\pm0.14$~mag in both $V$ and $I$ with a contribution to the
uncertainty of 0.13 from the distance estimate and 0.04 from the linear
fit. In $W$ the effect is very similar, $-0.21\pm0.14$ and in the $K$
band we can make the comparison with the Groenewegen (\cite{Groen00})
relation from Eq.(\ref{eq.logPMk_SMC}) where we find a very similar
effect of $\Delta M_K = -0.16\pm0.14$.

In the second approach we use the free fits (slope and zero-point) for the two
samples and perform the comparison at different periods. At $\log
P=1.18$, which is the average period of the SMC sample of Cepheids for
which we have determined surface brightness distances, we find values of
$\Delta M=-0.25$ for $V$ and $I$ with a similar uncertainty as above.
Due to the different slopes of the two relations the effect is less at
$\log P = 1.0$ and larger at $\log P = 1.3$. For completeness these values
have been included in Table \ref{tab.dM}. As in the first approach we have
compared the Galactic $K$~band relation with the Groenewegen (\cite{Groen00})
relation and the result is again similar to the one obtained for the
optical bands. 

  In the third approach we make a {\em direct} comparison between our
Baade-Wesselink sample stars and the Galactic sample independently of the
OGLE-2 sample. In this way we can investigate the metallicity effect in
all the bands available, including the reddening insensitive $K$~band,
without having to make any assumptions regarding a possible difference in
the slopes of the two P-L relations and we also avoid the application of
a correction for possible depth effects between the two samples.
For each SMC star we simply compute
the expected absolute magnitude from the Galactic relation. This value is
then compared with the absolute magnitude returned by the
Baade-Wesselink analysis of the star.  The offsets are
again given in Table \ref{tab.dM} where it can be seen that the effect
in $V$ and $I$ is smaller than found from the previous
two approaches by about 0.2~mag in $V$ and 0.12~mag in $I$ while the $K$
and $W$ offsets show very good agreement with the two previous methods.
This result adds further support to the size and direction of the
geometric depth correction which has been applied in
Sect.\ref{Sect.SMCdist}.

Having light curves in the $BVIJHK$~bands we can check the wavelength
dependence of the magnitude offsets. In Table \ref{tab.dMdFeHred} we give
the values for three different values for the assumed mean reddening. As
is expected, the values in the table show that any conclusion regarding
the wavelength dependence of the magnitude offsets is directly dependent
on the assumed reddening.  For an (unrealistic) reddening of $E(B-V)=0.0$
the offset is almost constant with wavelength, but as the reddening is
increased the metallicity sensitivity becomes a steeper and steeper
function of wavelength.

\begin{table}
\caption{\label{tab.dM}The magnitude offset between the Galactic and
SMC Cepheids for the different approaches mentioned in the text.}
\begin{tabular}{l c c c c}
\hline\hline
\multicolumn{1}{c}{Method} & $\Delta M_V$ & $\Delta M_I$ & $\Delta M_K$
& $\Delta M_W$ \\
                               & mag & mag & mag & mag \\
\hline
Fixed SMC slope, ZP fit only   & $-0.23$ & $-0.21$ & $-0.16$ & $-0.21$ \\
Free fits, for $\log P=1.0$    & $-0.16$ & $-0.16$ & $-0.09$ & $-0.16$ \\
Free fits, for $\log P=1.18$   & $-0.25$ & $-0.24$ & $-0.18$ & $-0.22$ \\
Free fits, for $\log P=1.3$    & $-0.31$ & $-0.29$ & $-0.23$ & $-0.26$ \\
Direct comparison of sample    & $-0.06$ & $-0.10$ & $-0.13$ & $-0.19$ \\
                               &         &         &         & \\
Best estimate (see text)       & $-0.15$ & $-0.16$ & $-0.15$ & $-0.20$\\
\hline
\end{tabular}
\end{table}

\begin{table}
\caption{\label{tab.dMdFeHred} The mean magnitude offsets from direct
comparison of the five $\log P \approx 1.18$ SMC stars with the Galactic
relations, for three different assumed values of the mean reddening.}
\begin{tabular}{c c c c}
\hline\hline
$E(B-V)$  & 0.00 & 0.06 & 0.12 \\
Band & $\Delta M$ & $\Delta M$ & $\Delta M$ \\
    &  mag    &   mag   &  mag    \\
\hline
$B$ & $-0.12$ & $+0.18$ & $+0.47$ \\
$V$ & $-0.24$ & $+0.00$ & $+0.24$ \\
$I$ & $-0.23$ & $-0.07$ & $+0.09$ \\
$J$ & $-0.21$ & $-0.13$ & $-0.04$ \\
$K$ & $-0.17$ & $-0.12$ & $-0.06$ \\
$W$ & $-0.22$ & $-0.18$ & $-0.14$ \\
\hline
\end{tabular}
\end{table}

  We note that the three methods described above all give the same
offsets in the $K$ and $W$ indices, and that the $K$ and $W$ offsets
agree well within the errors with each other. The $V$ and $I$~band
results are more dependent on accurate reddening estimates but give
basically the same offsets as the reddening independent indices, at
least for the first two methods. The third method suggests a smaller
effect in $V$ and $I$, but this is hardly significant.

 It should be stressed that we have direct measurements of the
offsets only at periods near $\log P = 1.18$.
At this period the two first methods are almost
equivalent and they can hardly be considered independent, which is also
reflected in the excellent agreement between the two sets of values. Our
best estimate for the offsets are thus computed as the weighted mean of
the first two approaches, each with weight 0.5, and the last approach
with a weight of unity. These estimates are also given in Table \ref{tab.dM}. 

Assuming a metallicity difference
between the Milky Way sample and the SMC sample of $0.7\pm0.07$~dex as
discussed in Sect.\ref{sec.SMC_FeH}, and adopting the individual
reddenings from Table \ref{tab.BVIred}, we have
measured the effect for 15 day period Cepheids to be $\Delta M_V /
\Delta \FeH = -0.21\pm0.19$, $\Delta M_I / \Delta \FeH = -0.23\pm 0.19$,
$\Delta M_W / \Delta \FeH = -0.29\pm0.19$, and
$\Delta M_K / \Delta \FeH = -0.21\pm 0.12$ where the minus sign
indicates that the absolute magnitude of a more metal-poor star is {\em
fainter} than a metal-rich counterpart of equal period. 
The error estimates are 1 $\sigma$ statistical errors which have
been propagated from the SMC distance error estimate and the metallicity error
estimate. The metallicity effect is
barely significant but in good agreement with the value adopted by
Freedman et al. (\cite{Freedman01}).

\subsection{The distance to the LMC}

Fouqu\'e et al. (\cite{FSG03})
have derived P-L relations for the OGLE-2 stars in the LMC. They
have forced the reddening to a common value of $E(B-V)=0.10$ to be in
agreement with the Hubble key project. They did not apply any correction
for metallicity but used the LMC slope of the P-L relations as the
slope for the Galactic Cepheids and simply fitted the zero-point for the
Galactic Cepheids on this basis. We follow the same procedure here and
assume that the slopes can be considered equal and apply our metallicity
corrections to the Fouqu\'e et al. (\cite{FSG03}) distances.
We assume that the metallicity effect is linear with $\FeH$ and simply apply the
metallicity effect for each index as determined in  the previous section
to the LMC moduli, under the assumption that 
$\FeH_{\mbox{\scriptsize LMC}} = -0.3$.  The corrected distance moduli in
the different bands can be found in Table \ref{tab.mMLMC} where our error
estimate includes a contribution from the uncertainty in the metallicity
correction of $0.3 \times 0.19=0.057$ added in quadrature.

\begin{table}
\caption{\label{tab.mMLMC} The distance modulus to the LMC in different
bands before (Fouqu\'e et al. \cite{FSG03}) and after
correction for metallicity.}
\begin{tabular}{r c c}
\hline\hline
Band & $(m-M)_0$ & $(m-M)_{0,c}$ \\
\hline
$V$ & $18.536\pm0.048$ & $18.47\pm0.07$ \\
$I$ & $18.530\pm0.041$ & $18.45\pm0.07$ \\
$K$ & $18.567\pm0.048$ & $18.50\pm0.07$ \\
$W$ & $18.549\pm0.036$ & $18.46\pm0.07$ \\
\hline
\end{tabular}
\end{table}

The values agree well within the stated errors. The $W$-index already combines
the $V$ and $I$ data in an optimum way, so to obtain a distance modulus free of
reddening uncertainties we simply compute a weighted mean of the $W$ and $K$
results, which leads to a best estimate of the LMC distance modulus of
$(m-M)_{0,c}=18.48\pm 0.07$. The error estimate is a formal error which
does not take into account possible systematic effects like differences
in the slopes between the Galactic and LMC Cepheid P-L relations.

We note that the metallicity corrected distance modulus is in good
agreement with the ISB distance of $(m-M)_0=18.42\pm0.10$ 
to the Cepheid HV12198 in the LMC cluster NGC1866 from 
Gieren et al. (\cite{Gieren00}) and is also in excellent agreement with
the canonical value of $18.50\pm0.1$ adopted by Freedman et al.
(\cite{Freedman01}).

\section{Discussion}

There have been several attempts to determine the metallicity
effect on the Cepheid P-L relation empirically, but it has proven
exceedingly difficult to reach truly conclusive results. 

Freedman \& Madore (\cite{FM90}) observed three fields at different
galactocentric distances in \object{M31}, and thus at different
metallicities. They found that the effect, if at all present, was
not significant.  This analysis and the results were later disputed by Gould
(\cite{Gould94}) who used a more sophisticated statistical treatment
of the data and found a strong effect ($\Delta \mu / \Delta \FeH =
-0.88\pm0.16$ \magdex).  Kennicutt et al. (\cite{Kenn98})
rediscussed these analyses pointing out that the assumed metallicity
range was incorrect, but even then they had to conclude that this data set
was not conclusive. Beaulieu et al. (\cite{Beaulieu97}) and Sasselov
et al. (\cite{Sasselov97}) applied a multi-variate analysis to the EROS
sample of SMC and LMC Cepheids to determine an effect which is equivalent
to $\Delta M_W / \Delta \FeH = -0.44^{+0.1}_{-0.2}$~\magdex. Kochanek
(\cite{Kochanek97}) used a similar technique for a sample of 17 galaxies
and Kennicutt et al. (\cite{Kenn98}) summarize the results as leading to
$\gamma_{VI}=\delta \mu_0/\delta [ O / H]=-0.4\pm0.2$~\magdex.  Kennicutt et al.
(\cite{Kenn98}) used a technique similar to the differential approach
of Freedman \& Madore (\cite{FM90}) in the galaxy \object{M101} and
found $\gamma_{VI}=-0.24\pm0.15$~\magdex.

Udalski et al. (\cite{Udalski01}) found a zero effect when comparing
the very low metallicity ($\FeH=-1.0$) galaxy IC1613 with the LMC.
This result depends critically on the assumed RR Lyrae
luminosity-metallicity relation which is itself a topic of debate.
As the galaxies have such low metallicity, the result is also not
necessarily representative of the solar-metallicity galaxies observed by the
Key Project, especially if the effect turns out not to be a linear
function of \FeH. If the effect happens to be linear in $Z$, the
heavy element fraction, then an extrapolation from the LMC to the Galaxy
would involve a factor of two as the Galactic Cepheids have a $Z$ twice
as big as that of the LMC Cepheids.

Freedman et al. (\cite{Freedman01})
have summarized the empirical evidence and assumed that the slope
of the P-L relation is universal and finally adopted a value
of $\gamma_{VI}=-0.2\pm0.2$~\magdex. 


In the light of our present results, and those discussed in Fouqu\'e et
al. (\cite{FSG03}), we 
share the concern of Tammann et al. (\cite{Tammann03}) that the slope
can be significantly different among populations. We have determined
our offsets at a mean period of $\log P = 1.18$ (15~days) which is not
the same as the mean period of $\log P \approx 1.5$ of most Key Project
Cepheid samples. If the slopes indeed do differ by as much as suggested
by our previous analysis, then our offsets would be larger by
approximately 0.1~mag when applied to samples of long ($\log P = 1.5$)
period Cepheids.

When we compare our offset in the $W$-index with that from the Key Project, 
we find that our values are in excellent agreement well within the
(admittedly large) errors. Including the slope variation and
performing the comparison at $\log P = 1.5$ would
make the difference larger. In $W$ we would find an offset between the
Galactic and SMC samples of 0.32~mag which
would result in $\Delta W / \Delta \FeH = -0.46\pm 0.19$, which is very
large, but still within the errors.

  On the theoretical side extensive efforts over the last few years by
several groups have not yet reached a consensus, and the situation is
maybe even more confusing than from the empirical point of view.

Bono et al. (\cite{Bono99}) have developed sequences of full amplitude 
non-linear, convective models for a large range of parameters. These
models have been further developed and analyzed in a number of papers.
In the latest paper in the series, Fiorentino et al.
(\cite{Fiorentino02}), it is argued that not only are Cepheid 
luminosities affected by metallicity, but also the Helium content
plays a significant role. The metallicity effect
is furthermore not a simple linear function in metallicity and even
the sign of the effect changes with abundance. 
These findings suggest that the problem might well be
significantly more complicated than the simple linear relation which has
so far been assumed for the empirical investigations.
Unfortunately these
computations suggest that metal-poor Cepheids are brighter than
metal-rich Cepheids for the metallicities and periods considered in the
present work. 

An alternative theoretical approach using linear non-adiabatic models
has been developed by Saio \& Gautschy (\cite{Saio98}), by Alibert et
al. (\cite{Alibert99}) and by Baraffe \& Alibert (\cite{Baraffe01}).
The first authors find a negligible metallicity
effect on the P-L relation while the latter authors find a small
effect of about 0.1~mag for the $V,I,J$, and $K$~bands in the sense
that metal-rich Cepheids are brighter than metal-poor ones.
This is in good agreement with our results. 

  Sandage et al. (\cite{Sandage99}) find similarly, from
a theoretical approach, that Cepheids with periods
between 10 and 30 days show a mild metallicity effect. They estimate the
effect to be $\Delta B / \Delta \FeH = +0.03$\magdex, 
$\Delta V / \Delta \FeH = -0.08$\magdex, and
$\Delta I / \Delta \FeH = -0.10$\magdex such that metal-rich stars are
brighter in the $V$ and $I$ bands, similarly to what we find.

It is reassuring that most of the independent empirical methods
lead to similar results, but the error bars are still too large to
distinguish among the available theoretical models. 
The sign of the effect for Cepheids of period
longer than about 10~days seems now to be reasonably well established,
at least from an empirical point of view, especially for the reddening
insensitive $K$- and $W$-indices. In addition to the large
empirical uncertainties it is also a serious concern that there is a
distinct possibility of a difference in P-L slope between the populations.
If such a difference is indeed present, the direct application of
the Galactic P-L relation presented here to the metal-rich
extra-galactic Cepheid samples and the LMC P-L relation 
to the metal-poor samples, would allow this potential problem to be
circumvented altogether as the metallicity difference between the
samples would be small.
Such an approach will depend on the adopted reddenings for
the Galactic stars, but this effect will be small in the reddening
insensitive $W$ and $K$ P-L relations.

  To place tighter constraints on the metallicity effect on both the
zero point and the slope of the P-L relation it would be very worthwhile
to extend this work to longer period Cepheids (35-50 days) in the SMC.
This would allow a direct comparison of the slopes of the P-L relations
for the two different metallicity populations using the same method. It
would also allow a comparison of the absolute magnitudes at a period
which better resemble the mean period of the Key Project sample which
becomes particularly important if a significant slope difference is
indeed present.

\section{Summary}

  We have presented Baade-Wesselink based absolute magnitudes for a
sample of 34 Galactic Cepheids and derived P-L relations spanning the period
range $\log P =0.6$ to $1.62$. For samples of solar-metallicity Cepheids in
other galaxies these relations allow a direct distance determination
without the need for an assumed LMC distance and without the need for a
significant correction for metallicity.

  We find some evidence for a difference in slope of the P-L relations
for the OGLE SMC sample and our Galactic Cepheid sample suggesting that
the metallicity effect might also affect the slopes of the P-L relations.

  Using stellar atmosphere models we have investigated the metallicity
effect on the infrared surface brightness method and we find only a
negligible effect. On this basis we find, on a purely differential basis,
that our sample of SMC Cepheids with periods around 15~days are
intrinsically slightly fainter than Galactic Cepheids with similar periods.

  Assuming that this effect is caused by the difference in metallicity
between
the two samples and assuming that this relation is linear over the range
of metallicities in question we find for the reddening insensitive
indices that 
$\Delta M_K / \Delta \FeH = -0.21\pm0.19$ and 
$\Delta M_W / \Delta \FeH = -0.29\pm0.19$. For our adopted reddening
towards the SMC Cepheids we find a similar effect for the
reddening sensitive indices, namely $\Delta M_V = -0.21\pm0.19$ and
$\Delta M_I = -0.23\pm0.19$. 
We consider the adopted reddening a best
estimate but realize that, if anything, the reddening could be stronger. 
This would cause the metallicity effect in the reddening sensitive
bands to be even closer to zero.

  We have determined the distance to the SMC based on the ISB distances
to individual Cepheids and by correcting for depth effects using a
geometric model. We find $(m-M)_0 = 18.88\pm0.13$.

  Fouqu\'e et al. (\cite{FSG03}) have determined the distance to the LMC
by applying the Galactic P-L relations from above to the OGLE data
(Udalski et al. \cite{Udalski99a}). We have corrected this distance
estimate for the effect of metallicity and find a final best estimate of
$(m-M)_0=18.48\pm0.07$ (internal error only).

  We finally conclude that our results support the two fundamental
assumptions of the Hubble Telescope Key Project on the Extragalactic
Distance scale, namely that the LMC distance modulus is $18.50\pm0.10$
and that the metallicity effect in the $W_{VI}$ index based distance is
$\gamma_{VI} = -0.2\pm0.2$~\magdex. However, if there is a significant
metallicity effect on the slope of the P-L relation, we would find a
stronger metallicity effect and a shorter LMC distance. 
In this case a direct application of the reddening insensitive Galactic 
P-L relations to the extra-galactic Cepheids would provide an attractive
alternative to using the LMC as an intermediate stepping stone.

\section*{Acknowledgements}
This research has made use of the SIMBAD database, operated at CDS,
Strasbourg, France, NASA's Astrophysics Data System 
Bibliographic Services, and the McMaster Cepheid Photometry and Radial
Velocity Data Archive. WPG acknowledges support for this work from the
chilean FONDAP Center for Astrophysics 15010003. BWC thanks the U.S.
National Science Foundation for grants AST-8920742 and AST-9800427 to
the University of North Carolina for support for this research.


\appendix

\section{New radial velocities}

New measurements of the radial velocities for ten of the Galactic
cepheids have been obtained with the CfA Digital Speedometers
Latham (\cite{Lath85},\cite{Lath92}). 
Three nearly identical echelle spectrographs
have been used, on the Multiple Mirror Telescope and 1.5-m Tillinghast
Reflector at the F.\ L.\ Whipple Observatory atop Mt.\ Hopkins,
Arizona, and on the 1.5-m Wyeth Reflector located at the Oak Ridge
Observatory in the town of Harvard, Massachusetts. Photon-counting
intensified Reticon detectors were used to record about 45 \AA\ of
spectrum in a single order centered near 5187 \AA.  The spectral
resolution is about 8.5 \kms\ for all the CfA spectra, and the
signal-to-noise ratios range from about 5 to 50 per resolution
element.

\begin{table}
\caption{\label{tab.templates} The best fitting
template parameters, the number of observations, $N_{\mbox{\scriptsize
obs}}$, and the time span of the observations. The
last two columns give the rotational velocity and the systemic velocity.
}
\begin{tabular}{r c c c c r r}
\hline\hline
ID & \Teff & $\log(g)$ & $N_{\mbox{\scriptsize obs}}$ & Span &
\multicolumn{1}{c}{$V_{\mbox{\scriptsize rot}}$} &
\multicolumn{1}{c}{$\gamma$} \\
   &  K    &           &                            & days &
\kms                       & \kms \\
\hline
    \object{SU Cas} &  6500 &  1.5 &   53 &  944 &   15.0 &  $-7.3$ \\
    \object{EV Sct} &  6250 &  1.5 &   67 &  890 &   22.3 &  $17.3$ \\
\object{$\delta$ Cep} &  5750 &  1.5 &   20 &  126 &   15.2 & $-16.3$ \\
    \object{CV Mon} &  6000 &  1.5 &   64 &  862 &   17.1 &  $19.4$ \\
     \object{U Sgr} &  5750 &  1.5 &   57 &  749 &   19.7 &   $2.8$ \\
\object{$\eta$ Aql} &  5750 &  1.5 &   26 &  467 &   18.3 & $-15.5$ \\
     \object{X Cyg} &  5250 &  0.5 &   75 &  806 &   21.2 &  $ 7.5$ \\
     \object{T Mon} &  5250 &  0.5 &   88 &  937 &   21.9 &  $20.3$ \\
    \object{RS Pup} &  5000 &  0.5 &   43 &  718 &   20.2 &  $24.7$ \\
    \object{SV Vul} &  5250 &  0.5 &   75 &  839 &   22.4 &  $-1.3$ \\
\hline
\end{tabular}
\end{table}

Radial velocities were extracted from the observed spectra using the
one-dimensional correlation package {\tt r2rvsao} 
(Kurtz \& Mink \cite{KurtM98})
running inside the IRAF\footnotemark[1] \footnotetext[1]{IRAF (Image
Reduction and Analysis Facility) is distributed by the National
Optical Astronomy Observatories, which are operated by the Association
of Universities for Research in Astronomy, Inc., under contract with
the National Science Foundation.}
environment.  For the templates we used a new grid of synthetic
spectra (Morse \& Kurucz, priv.  comm.)
calculated using the latest Kurucz model atmospheres.  The new grid of
synthetic templates incorporates several improvements compared to the
older grid that we used for several years (Nordstr\"om et al.
\cite{Noretal94}, Latham et al. \cite{Latetal02}).
To select the optimum synthetic template for each star we adopted
solar metallicity and ran correlations for an appropriate range of
effective temperature, \Teff, surface gravity, \logg, and
rotational velocity.  The \Teff and \logg for the templates
that gave the highest average value for the peak correlation
coefficient for the individual stars are listed in Table
\ref{tab.templates}. Of course, both \Teff and \logg vary
systematically during the pulsational cycle of a Cepheid, so the values
that we adopted for the template parameters can only represent some
kind of average approximation. Similarly, the width of the observed
correlation peaks vary significantly as a function of pulsational
phase.  One obvious source of variation in the observed line
broadening with phase is the change in the pulsational velocity and its
projection as one moves from the center to the edge of the disk.  A
more sophisticated approach to the analysis of cepheid velocities
would be to match calculated spectra to each individual velocity and
thereby derive \Teff and \logg at each observed phase in
the pulsational cycle.  One can even imagine an approach that would
add an extra dimension to the grid of synthetic template spectra,
namely the pulsational velocity, taking into account explicitly the
effect of changing projection factor across the disk.

The template parameters derived from the CfA spectra follow the
expected trend in the sense that the cepheids with longer periods are
also cooler and have weaker surface gravity. Although it is reassuring
that our procedure picks stellar parameters that appear to have some
connection with reality, we must caution that our procedure was
designed to optimize the radial-velocity determinations, not to
determine the fundamental astrophysical characteristics of our stars.
In particular, if the actual metallicity of a star is significantly
different from solar, the \Teff and \logg values that give
the best correlations will exhibit systematic errors. Furthermore, the
most luminous cepheids in our sample give the best correlations for
synthetic spectra at the edge of our grid at $\logg = 0.5$.

\begin{figure*}[t]
\epsfig{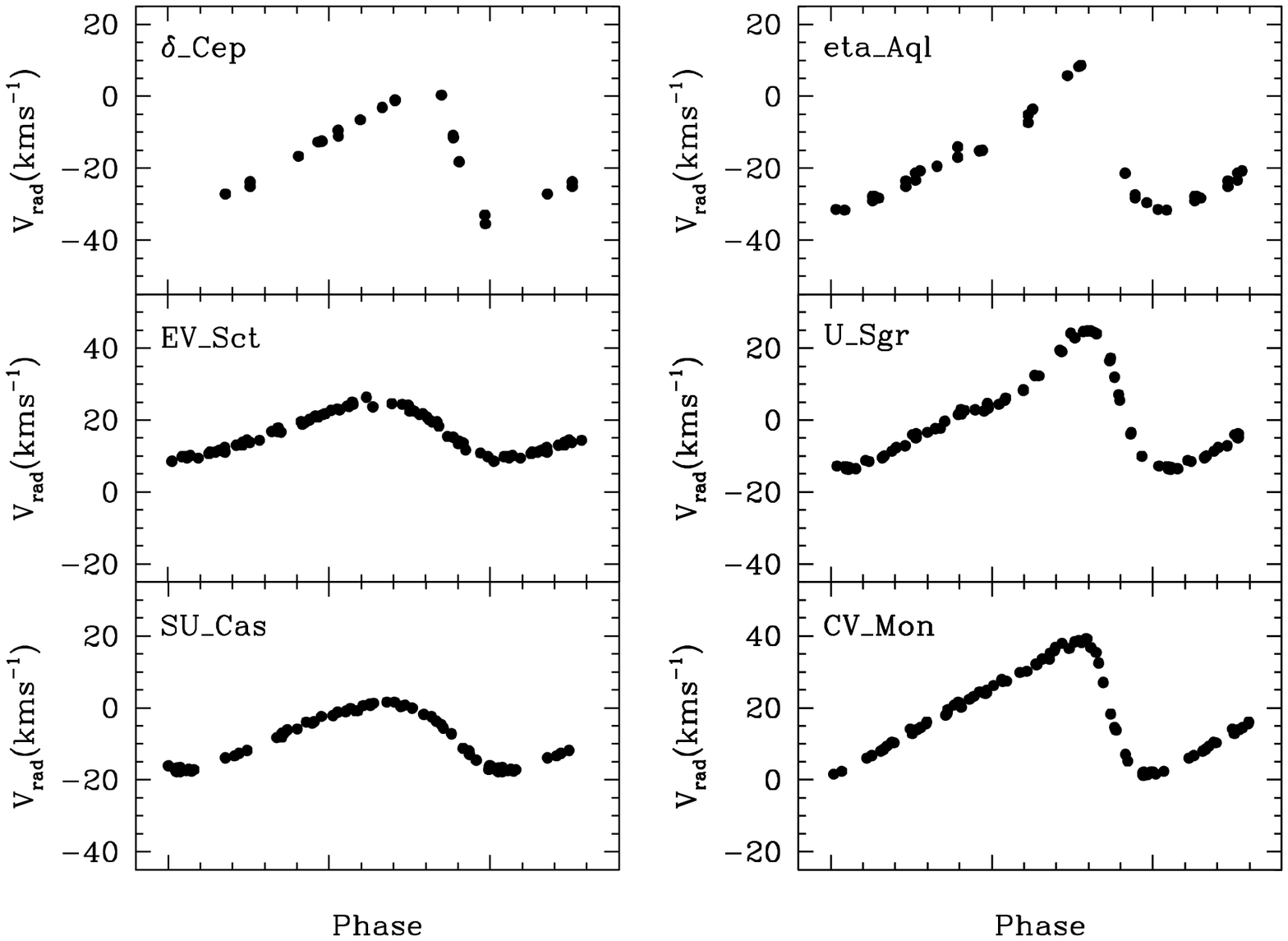}
\figcap{\label{fig.CfArvall1} The radial velocity curves for the stars
tabulated in Table \ref{tab.rvSUCas}-\ref{tab.rvetaAql}.}
\end{figure*}

\begin{figure*}[htp]
\epsfig{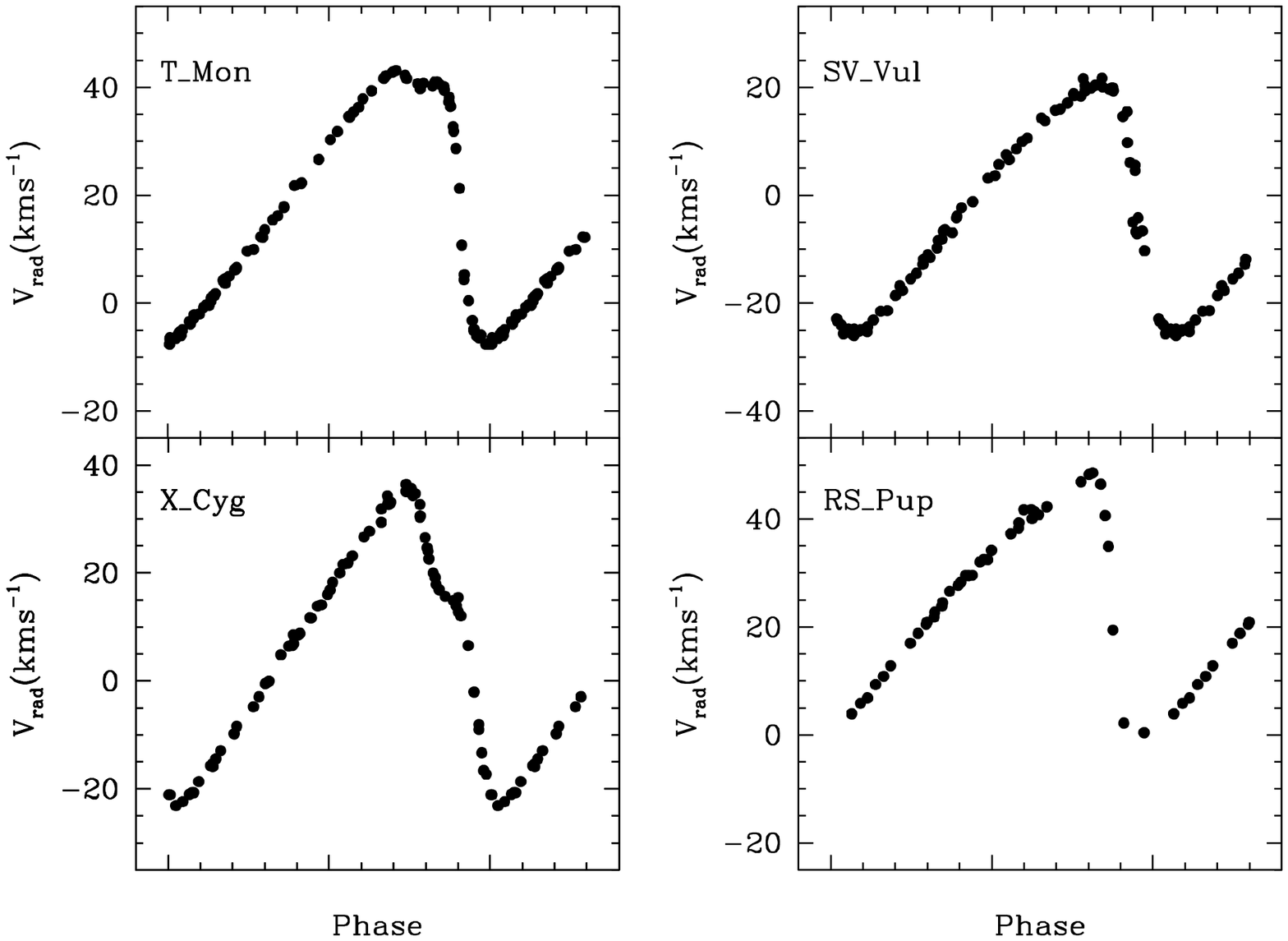}
\figcap{\label{fig.CfArvall2} The radial velocity curves for the stars
tabulated in Table \ref{tab.rvXCyg}-\ref{tab.rvSVVul}.}
\end{figure*}

  The resulting radial-velocity data is tabulated for each star in
Table \ref{tab.rvSUCas}-\ref{tab.rvSVVul}
and the corresponding radial-velocity curves
are plotted in Fig.\ref{fig.CfArvall1}-\ref{fig.CfArvall2}. 
Here the excellent
coverage for most of the stars is apparent as well as the low scatter.
Note that the estimated errors are typically less than 0.5~\kms.

%
%
%

{}

\clearpage

\begin{table*}
\caption{\label{tab.rvSUCas} Radial velocities
for \object{SU~Cas} tabulated with heliocentric Julian date (HJD), phase
and telescope identifier. "T" referes to the Tillinghast reflector, "W"
to the Wyeth reflector and "M" to the Multiple Mirror Telescope.}
\small
\begin{tabular}{c c r c c| c c r c c| c c r c c}
\hline\hline
HJD & phase & \multicolumn{1}{c}{$V_{\mbox{\scriptsize rad}}$} &
\multicolumn{1}{c}{$\sigma$} & Tel.&
HJD & phase & \multicolumn{1}{c}{$V_{\mbox{\scriptsize rad}}$} &
\multicolumn{1}{c}{$\sigma$} & Tel.&
HJD & phase & \multicolumn{1}{c}{$V_{\mbox{\scriptsize rad}}$} &
\multicolumn{1}{c}{$\sigma$} & Tel.\\
$-2400000$ & & & & &
$-2400000$ & & & & &
$-2400000$ & & & & \\
days & & \multicolumn{1}{c}{\kms} & \multicolumn{1}{c}{\kms} & &
days & & \multicolumn{1}{c}{\kms} & \multicolumn{1}{c}{\kms} & &
days & & \multicolumn{1}{c}{\kms}  & \multicolumn{1}{c}{\kms} & \\
\hline
46253.9516 & 0.916 & $ -11.2$ & 0.3 &  M & 46787.7007 & 0.728 & $   0.5$ & 0.3 &  T & 47107.9642 & 0.023 & $ -17.5$ & 0.4 &  T \\
46254.8979 & 0.401 & $  -5.8$ & 0.3 &  M & 46814.7194 & 0.589 & $  -0.8$ & 0.3 &  T & 47109.7385 & 0.934 & $ -12.0$ & 0.4 &  T \\
46254.9525 & 0.429 & $  -3.9$ & 0.4 &  M & 46815.5937 & 0.038 & $ -17.8$ & 0.4 &  T & 47109.8694 & 0.001 & $ -16.1$ & 0.4 &  T \\
46721.8685 & 0.957 & $ -14.5$ & 0.3 &  T & 46816.7021 & 0.606 & $   0.7$ & 0.2 &  T & 47135.8967 & 0.353 & $  -7.0$ & 0.3 &  T \\
46727.7935 & 0.996 & $ -17.2$ & 0.3 &  T & 46864.6553 & 0.206 & $ -13.4$ & 0.3 &  T & 47137.8157 & 0.337 & $  -8.3$ & 0.3 &  T \\
46728.7295 & 0.476 & $  -2.4$ & 0.3 &  T & 46865.6606 & 0.722 & $   0.4$ & 0.3 &  T & 47138.7527 & 0.818 & $  -2.5$ & 0.3 &  T \\
46729.8590 & 0.056 & $ -17.4$ & 0.3 &  T & 46866.6796 & 0.245 & $ -11.8$ & 0.2 &  T & 47140.7313 & 0.833 & $  -3.7$ & 0.3 &  T \\
46730.9945 & 0.638 & $   1.3$ & 0.3 &  T & 46867.6804 & 0.758 & $  -0.1$ & 0.2 &  T & 47141.7781 & 0.370 & $  -6.1$ & 0.3 &  T \\
46754.7953 & 0.848 & $  -4.6$ & 0.3 &  T & 46868.5819 & 0.220 & $ -12.7$ & 0.3 &  T & 47158.6527 & 0.026 & $ -17.7$ & 0.4 &  T \\
46755.7802 & 0.353 & $  -8.1$ & 0.3 &  T & 46869.5859 & 0.736 & $   0.8$ & 0.3 &  T & 47159.5995 & 0.512 & $  -2.1$ & 0.3 &  T \\
46756.7576 & 0.855 & $  -5.6$ & 0.3 &  T & 46873.5981 & 0.794 & $  -1.8$ & 0.3 &  T & 47160.6015 & 0.026 & $ -16.7$ & 0.3 &  T \\
46757.7446 & 0.361 & $  -6.9$ & 0.2 &  T & 46891.6153 & 0.037 & $ -16.5$ & 0.4 &  T & 47165.5836 & 0.582 & $  -0.8$ & 0.3 &  T \\
46773.7141 & 0.553 & $  -0.9$ & 0.3 &  T & 46894.5960 & 0.566 & $  -0.1$ & 0.2 &  T & 47171.6671 & 0.703 & $   1.6$ & 0.3 &  T \\
46774.7273 & 0.073 & $ -17.6$ & 0.3 &  T & 46895.5977 & 0.080 & $ -17.2$ & 0.3 &  T & 47191.6139 & 0.936 & $ -12.9$ & 0.3 &  T \\
46777.6075 & 0.551 & $  -1.1$ & 0.3 &  T & 46901.6379 & 0.178 & $ -13.9$ & 0.3 &  T & 47192.6229 & 0.453 & $  -3.9$ & 0.3 &  T \\
46778.6086 & 0.064 & $ -17.0$ & 0.3 &  T & 46902.6151 & 0.679 & $   1.6$ & 0.3 &  T & 47197.5826 & 0.997 & $ -16.1$ & 0.4 &  T \\
46779.7046 & 0.626 & $   1.1$ & 0.3 &  T & 47106.8405 & 0.447 & $  -4.2$ & 0.3 &  T & 47198.6124 & 0.526 & $  -1.2$ & 0.3 &  T \\
46783.6070 & 0.628 & $   0.8$ & 0.3 &  T & 47107.6846 & 0.880 & $  -7.2$ & 0.4 &  T &  &  &  &  &  \\
\hline
\end{tabular}
\end{table*}

\begin{table*}
\caption{\label{tab.rvEVSct} Radial velocities
for \object{EV~Sct}
tabulated with heliocentric Julian date, phase and telescope identifier
as in Tab.\ref{tab.rvSUCas}}
\small
\begin{tabular}{c c r r c| c c r r c| c c r r c}
\hline\hline
HJD & phase & \multicolumn{1}{c}{$V_{\mbox{\scriptsize rad}}$} &
$\sigma$ & Tel.&
HJD & phase & \multicolumn{1}{c}{$V_{\mbox{\scriptsize rad}}$} &
$\sigma$ & Tel.&
HJD & phase & \multicolumn{1}{c}{$V_{\mbox{\scriptsize rad}}$} &
$\sigma$ & Tel.\\
$-2400000$ & & & & &
$-2400000$ & & & & &
$-2400000$ & & & & \\
days & & \multicolumn{1}{c}{\kms} & \multicolumn{1}{c}{\kms} & &
days & & \multicolumn{1}{c}{\kms} & \multicolumn{1}{c}{\kms} & &
days & & \multicolumn{1}{c}{\kms}  & \multicolumn{1}{c}{\kms} & \\
\hline
46567.8871 & 0.758 & $  22.8$ & 0.6 &  T & 46932.9633 & 0.868 & $  15.4$ & 0.6 &  T & 47281.9760 & 0.781 & $  21.5$ & 0.5 &  T \\
46569.9099 & 0.413 & $  19.6$ & 0.4 &  T & 46933.9078 & 0.174 & $  12.3$ & 0.5 &  T & 47283.9675 & 0.425 & $  19.3$ & 0.5 &  T \\
46571.9060 & 0.059 & $   9.4$ & 0.5 &  T & 46956.9125 & 0.616 & $  26.3$ & 0.8 &  T & 47285.9570 & 0.069 & $  10.1$ & 0.7 &  T \\
46577.8843 & 0.993 & $   9.8$ & 0.5 &  T & 46957.8652 & 0.924 & $  11.7$ & 0.8 &  T & 47306.8693 & 0.835 & $  19.5$ & 0.6 &  T \\
46578.9628 & 0.342 & $  17.9$ & 0.6 &  T & 46958.8552 & 0.245 & $  14.5$ & 0.5 &  T & 47307.9273 & 0.177 & $  11.1$ & 0.5 &  T \\
46595.8083 & 0.791 & $  21.8$ & 0.7 &  T & 46959.8161 & 0.556 & $  23.9$ & 0.6 &  T & 47308.8879 & 0.488 & $  21.8$ & 0.6 &  T \\
46595.8479 & 0.804 & $  20.8$ & 0.6 &  T & 46960.8905 & 0.903 & $  14.1$ & 0.5 &  T & 47309.9789 & 0.841 & $  18.3$ & 0.5 &  T \\
46596.8541 & 0.130 & $  11.2$ & 0.3 &  T & 46979.7735 & 0.012 & $   8.5$ & 0.7 &  T & 47310.9650 & 0.160 & $  11.7$ & 0.4 &  T \\
46596.9059 & 0.147 & $  11.1$ & 0.5 &  T & 46980.7856 & 0.340 & $  16.8$ & 0.6 &  T & 47314.9193 & 0.439 & $  20.0$ & 0.7 &  T \\
46597.8890 & 0.465 & $  20.8$ & 0.5 &  T & 46983.8215 & 0.322 & $  16.8$ & 0.5 &  T & 47318.9599 & 0.746 & $  24.1$ & 0.7 &  T \\
46597.9507 & 0.485 & $  21.7$ & 0.5 &  T & 46984.7925 & 0.636 & $  23.7$ & 0.5 &  T & 47319.8769 & 0.043 & $   9.8$ & 0.7 &  T \\
46598.8155 & 0.764 & $  22.5$ & 0.5 &  T & 46985.8239 & 0.970 & $  10.8$ & 0.6 &  T & 47335.8541 & 0.212 & $  13.0$ & 0.4 &  T \\
46600.8317 & 0.417 & $  18.8$ & 0.7 &  T & 46986.7944 & 0.284 & $  14.4$ & 0.5 &  T & 47336.8477 & 0.533 & $  22.9$ & 0.4 &  T \\
46601.7903 & 0.727 & $  24.4$ & 0.6 &  T & 46989.7925 & 0.254 & $  13.8$ & 0.5 &  T & 47337.9370 & 0.886 & $  15.2$ & 0.7 &  T \\
46601.8612 & 0.750 & $  22.5$ & 0.4 &  T & 46990.7462 & 0.562 & $  23.8$ & 0.6 &  T & 47338.8351 & 0.176 & $  12.4$ & 0.5 &  T \\
46626.8050 & 0.820 & $  19.5$ & 0.7 &  T & 46990.7872 & 0.575 & $  24.4$ & 0.5 &  M & 47339.8542 & 0.506 & $  22.7$ & 0.6 &  T \\
46638.7835 & 0.695 & $  24.6$ & 0.4 &  T & 46993.8664 & 0.572 & $  25.0$ & 0.7 &  M & 47340.7988 & 0.811 & $  20.1$ & 0.6 &  T \\
46693.6874 & 0.457 & $  21.2$ & 0.5 &  T & 46995.6970 & 0.164 & $  11.6$ & 0.5 &  M & 47344.7617 & 0.094 & $   9.4$ & 0.6 &  T \\
46694.6188 & 0.759 & $  22.6$ & 0.5 &  T & 47053.6598 & 0.916 & $  13.7$ & 0.5 &  T & 47369.7371 & 0.174 & $  11.2$ & 0.6 &  T \\
46721.5757 & 0.480 & $  21.5$ & 0.6 &  T & 47054.6334 & 0.231 & $  13.1$ & 0.5 &  T & 47437.5957 & 0.127 & $  10.7$ & 0.5 &  T \\
46900.9916 & 0.525 & $  23.1$ & 0.5 &  T & 47072.6295 & 0.053 & $   9.8$ & 0.6 &  T & 47458.5546 & 0.908 & $  13.8$ & 0.6 &  T \\
46929.9745 & 0.901 & $  13.4$ & 0.6 &  T & 47076.6402 & 0.351 & $  16.7$ & 0.7 &  T &  &  &  &  &  \\
46930.9930 & 0.231 & $  13.8$ & 0.6 &  T & 47280.9192 & 0.439 & $  20.2$ & 0.5 &  T &  &  &  &  &  \\
\hline
\end{tabular}
\end{table*}

\begin{table*}
\caption{\label{tab.rvdCeph} Radial velocities
for \object{$\delta$~Cep}
tabulated with heliocentric Julian date, phase and telescope identifier
as in Tab.\ref{tab.rvSUCas}}
\small
\begin{tabular}{c c r r c| c c r r c| c c r r c}
\hline\hline
HJD & phase & \multicolumn{1}{c}{$V_{\mbox{\scriptsize rad}}$} &
$\sigma$ & Tel.&
HJD & phase & \multicolumn{1}{c}{$V_{\mbox{\scriptsize rad}}$} &
$\sigma$ & Tel.&
HJD & phase & \multicolumn{1}{c}{$V_{\mbox{\scriptsize rad}}$} &
$\sigma$ & Tel.\\
$-2400000$ & & & & &
$-2400000$ & & & & &
$-2400000$ & & & & \\
days & & \multicolumn{1}{c}{\kms} & \multicolumn{1}{c}{\kms} & &
days & & \multicolumn{1}{c}{\kms} & \multicolumn{1}{c}{\kms} & &
days & & \multicolumn{1}{c}{\kms}  & \multicolumn{1}{c}{\kms} & \\
\hline
46595.8671 & 0.466 & $ -12.7$ & 0.3 &  T & 46630.8430 & 0.984 & $ -33.0$ & 0.7 &  T & 46692.7939 & 0.529 & $ -11.1$ & 0.3 &  T \\
46596.9338 & 0.665 & $  -3.1$ & 0.3 &  T & 46630.8455 & 0.985 & $ -35.5$ & 0.4 &  T & 46693.7403 & 0.705 & $  -1.2$ & 0.4 &  T \\
46597.9195 & 0.849 & $   0.3$ & 0.4 &  T & 46631.8817 & 0.178 & $ -27.1$ & 0.5 &  T & 46693.7425 & 0.705 & $  -1.0$ & 0.3 &  T \\
46600.9019 & 0.405 & $ -16.6$ & 0.3 &  T & 46631.8841 & 0.178 & $ -27.2$ & 0.5 &  T & 46694.7062 & 0.885 & $ -10.8$ & 0.4 &  T \\
46601.9334 & 0.597 & $  -6.5$ & 0.3 &  T & 46638.8551 & 0.477 & $ -12.5$ & 0.3 &  T & 46694.7099 & 0.886 & $ -11.5$ & 0.4 &  T \\
46626.9301 & 0.255 & $ -25.1$ & 0.4 &  T & 46638.8587 & 0.478 & $ -12.5$ & 0.3 &  T & 46721.6390 & 0.904 & $ -18.2$ & 0.4 &  T \\
46626.9326 & 0.255 & $ -23.8$ & 0.4 &  T & 46692.7902 & 0.528 & $  -9.5$ & 0.3 &  T &  &  &  &  &  \\
\hline
\end{tabular}
\end{table*}

\begin{table*}
\caption{\label{tab.rvCVMon} Radial velocities
for \object{CV~Mon}
tabulated with heliocentric Julian date, phase and telescope identifier
as in Tab.\ref{tab.rvSUCas}}
\small
\begin{tabular}{c c r r c| c c r r c| c c r r c}
\hline\hline
HJD & phase & \multicolumn{1}{c}{$V_{\mbox{\scriptsize rad}}$} &
$\sigma$ & Tel.&
HJD & phase & \multicolumn{1}{c}{$V_{\mbox{\scriptsize rad}}$} &
$\sigma$ & Tel.&
HJD & phase & \multicolumn{1}{c}{$V_{\mbox{\scriptsize rad}}$} &
$\sigma$ & Tel.\\
$-2400000$ & & & & &
$-2400000$ & & & & &
$-2400000$ & & & & \\
days & & \multicolumn{1}{c}{\kms} & \multicolumn{1}{c}{\kms} & &
days & & \multicolumn{1}{c}{\kms} & \multicolumn{1}{c}{\kms} & &
days & & \multicolumn{1}{c}{\kms}  & \multicolumn{1}{c}{\kms} & \\
\hline
46722.0037 & 0.155 & $   8.0$ & 0.5 &  T & 46895.6807 & 0.444 & $  23.2$ & 0.4 &  T & 47224.6721 & 0.609 & $  30.2$ & 0.5 &  T \\
46727.9958 & 0.269 & $  14.0$ & 0.5 &  T & 46898.6300 & 0.993 & $   2.2$ & 0.5 &  T & 47225.6748 & 0.795 & $  39.2$ & 0.5 &  T \\
46729.0348 & 0.462 & $  24.5$ & 0.4 &  T & 46900.6201 & 0.363 & $  19.5$ & 0.3 &  T & 47226.6094 & 0.969 & $   1.4$ & 0.7 &  T \\
46729.9770 & 0.637 & $  32.2$ & 0.4 &  T & 47107.0445 & 0.740 & $  36.5$ & 0.5 &  T & 47227.6566 & 0.164 & $   8.5$ & 0.3 &  T \\
46731.0225 & 0.832 & $  32.5$ & 0.5 &  T & 47108.0248 & 0.922 & $   5.2$ & 0.5 &  T & 47229.6496 & 0.534 & $  27.4$ & 0.6 &  T \\
46757.8725 & 0.824 & $  35.4$ & 0.5 &  T & 47109.9412 & 0.279 & $  14.5$ & 0.4 &  T & 47230.6359 & 0.718 & $  37.9$ & 0.6 &  T \\
46773.8321 & 0.791 & $  39.1$ & 0.5 &  T & 47135.9379 & 0.112 & $   6.0$ & 0.5 &  T & 47250.6060 & 0.430 & $  22.4$ & 0.4 &  T \\
46774.8206 & 0.974 & $   1.3$ & 0.5 &  T & 47136.9289 & 0.296 & $  15.6$ & 0.5 &  T & 47251.7297 & 0.639 & $  32.1$ & 0.5 &  T \\
46779.8807 & 0.915 & $   7.1$ & 0.5 &  T & 47137.9326 & 0.483 & $  24.1$ & 0.6 &  T & 47252.6393 & 0.808 & $  36.7$ & 0.6 &  T \\
46782.9185 & 0.480 & $  24.0$ & 0.5 &  T & 47139.8851 & 0.846 & $  27.1$ & 0.6 &  T & 47253.6688 & 1.000 & $   2.2$ & 0.6 &  T \\
46783.8688 & 0.657 & $  33.6$ & 0.4 &  T & 47140.8982 & 0.034 & $   2.4$ & 0.4 &  T & 47254.6875 & 0.189 & $  10.4$ & 0.4 &  T \\
46787.8473 & 0.396 & $  21.6$ & 0.4 &  T & 47158.7709 & 0.357 & $  18.0$ & 0.4 &  T & 47255.6203 & 0.362 & $  18.6$ & 0.4 &  T \\
46814.7854 & 0.405 & $  20.2$ & 0.3 &  T & 47159.7864 & 0.545 & $  27.4$ & 0.4 &  T & 47278.6223 & 0.639 & $  31.9$ & 0.5 &  T \\
46815.7669 & 0.587 & $  29.8$ & 0.4 &  T & 47163.8315 & 0.298 & $  16.1$ & 0.6 &  T & 47280.6138 & 0.009 & $   1.6$ & 0.6 &  T \\
46816.7491 & 0.770 & $  38.7$ & 0.6 &  T & 47191.7276 & 0.484 & $  24.9$ & 0.6 &  T & 47281.6185 & 0.196 & $  10.3$ & 0.4 &  T \\
46817.8170 & 0.968 & $   1.9$ & 0.4 &  T & 47192.7936 & 0.682 & $  35.3$ & 0.4 &  T & 47282.6241 & 0.383 & $  20.8$ & 0.4 &  T \\
46844.7274 & 0.971 & $   2.1$ & 0.4 &  T & 47198.6914 & 0.779 & $  38.2$ & 0.4 &  T & 47284.6357 & 0.757 & $  38.3$ & 0.5 &  T \\
46864.6686 & 0.679 & $  33.6$ & 0.4 &  T & 47216.7029 & 0.127 & $   6.7$ & 0.4 &  T & 47457.9768 & 0.984 & $   1.5$ & 0.5 &  T \\
46865.6945 & 0.869 & $  18.3$ & 0.6 &  T & 47218.7367 & 0.505 & $  26.2$ & 0.7 &  T & 47459.0021 & 0.174 & $   9.4$ & 0.4 &  T \\
46891.6602 & 0.697 & $  36.8$ & 0.6 &  T & 47219.7557 & 0.695 & $  36.0$ & 0.6 &  T & 47584.6327 & 0.531 & $  27.9$ & 0.4 &  T \\
46892.6769 & 0.886 & $  13.8$ & 0.5 &  T & 47220.7556 & 0.881 & $  14.6$ & 0.7 &  T &  &  &  &  &  \\
46894.6541 & 0.253 & $  12.8$ & 0.4 &  T & 47222.7270 & 0.247 & $  14.1$ & 0.4 &  T &  &  &  &  &  \\
\hline
\end{tabular}
\end{table*}

\begin{table*}
\caption{\label{tab.rvUSgr} Radial velocities
for \object{U~Sgr}
tabulated with heliocentric Julian date, phase and telescope identifier
as in Tab.\ref{tab.rvSUCas}}
\small
\begin{tabular}{c c r r c| c c r r c| c c r r c}
\hline\hline
HJD & phase & \multicolumn{1}{c}{$V_{\mbox{\scriptsize rad}}$} &
$\sigma$ & Tel.&
HJD & phase & \multicolumn{1}{c}{$V_{\mbox{\scriptsize rad}}$} &
$\sigma$ & Tel.&
HJD & phase & \multicolumn{1}{c}{$V_{\mbox{\scriptsize rad}}$} &
$\sigma$ & Tel.\\
$-2400000$ & & & & &
$-2400000$ & & & & &
$-2400000$ & & & & \\
days & & \multicolumn{1}{c}{\kms} & \multicolumn{1}{c}{\kms} & &
days & & \multicolumn{1}{c}{\kms} & \multicolumn{1}{c}{\kms} & &
days & & \multicolumn{1}{c}{\kms}  & \multicolumn{1}{c}{\kms} & \\
\hline
46567.8965 & 0.758 & $  22.9$ & 0.5 &  T & 46714.4569 & 0.486 & $   4.6$ & 0.4 &  W & 46952.7787 & 0.818 & $  24.4$ & 0.6 &  W \\
46569.8999 & 0.055 & $ -13.7$ & 0.5 &  T & 46715.4523 & 0.633 & $  12.4$ & 0.5 &  W & 46963.7774 & 0.448 & $   2.9$ & 0.4 &  W \\
46571.9195 & 0.354 & $  -0.3$ & 0.4 &  T & 46719.4813 & 0.231 & $  -7.1$ & 0.4 &  W & 46964.7864 & 0.598 & $   8.4$ & 0.5 &  W \\
46578.9473 & 0.396 & $   1.5$ & 0.4 &  T & 46721.5691 & 0.540 & $   5.5$ & 0.5 &  T & 46982.7000 & 0.254 & $  -4.0$ & 0.4 &  W \\
46595.8161 & 0.897 & $   5.5$ & 0.6 &  T & 46723.4926 & 0.825 & $  24.0$ & 0.5 &  W & 46994.6082 & 0.019 & $ -12.8$ & 0.6 &  W \\
46596.8364 & 0.048 & $ -13.6$ & 0.5 &  T & 46724.4408 & 0.966 & $ -10.0$ & 0.6 &  W & 47020.5587 & 0.866 & $  16.5$ & 0.7 &  W \\
46596.8961 & 0.057 & $ -13.2$ & 0.5 &  T & 46726.4553 & 0.264 & $  -5.0$ & 0.4 &  W & 47020.6613 & 0.881 & $  11.9$ & 0.6 &  W \\
46597.8799 & 0.203 & $  -7.7$ & 0.4 &  T & 46736.4408 & 0.745 & $  24.1$ & 0.6 &  W & 47022.5411 & 0.160 & $ -10.5$ & 0.5 &  W \\
46598.8072 & 0.340 & $  -2.3$ & 0.5 &  T & 46739.4401 & 0.190 & $  -8.6$ & 0.4 &  W & 47022.5841 & 0.166 & $ -10.0$ & 0.5 &  W \\
46600.8712 & 0.646 & $  12.3$ & 0.4 &  T & 46745.4285 & 0.077 & $ -13.5$ & 0.6 &  W & 47023.6543 & 0.325 & $  -2.4$ & 0.5 &  W \\
46601.8020 & 0.784 & $  24.6$ & 0.5 &  T & 46748.4297 & 0.522 & $   4.4$ & 0.4 &  W & 47024.6730 & 0.476 & $   2.5$ & 0.6 &  W \\
46601.8832 & 0.796 & $  24.8$ & 0.5 &  T & 46760.4182 & 0.300 & $  -3.4$ & 0.4 &  W & 47042.5018 & 0.119 & $ -11.5$ & 0.4 &  W \\
46626.7892 & 0.489 & $   3.3$ & 0.6 &  T & 46927.7570 & 0.108 & $ -11.2$ & 0.7 &  W & 47082.4743 & 0.045 & $ -12.9$ & 0.5 &  W \\
46638.7695 & 0.265 & $  -4.9$ & 0.3 &  T & 46929.8257 & 0.415 & $   2.7$ & 0.4 &  W & 47085.4547 & 0.487 & $   4.4$ & 0.5 &  W \\
46638.7728 & 0.265 & $  -3.8$ & 0.2 &  T & 46931.8226 & 0.711 & $  19.4$ & 0.4 &  W & 47115.4269 & 0.931 & $  -3.9$ & 0.5 &  W \\
46693.6704 & 0.404 & $   3.0$ & 0.3 &  T & 46931.8677 & 0.717 & $  19.1$ & 0.5 &  W & 47222.9352 & 0.869 & $  17.2$ & 0.9 &  W \\
46693.6760 & 0.405 & $   1.6$ & 0.3 &  T & 46937.8043 & 0.598 & $   8.2$ & 0.5 &  W & 47310.7938 & 0.894 & $   7.1$ & 0.6 &  W \\
46694.6108 & 0.543 & $   6.1$ & 0.3 &  T & 46952.7064 & 0.807 & $  24.8$ & 0.6 &  W & 47317.7979 & 0.933 & $  -3.5$ & 0.6 &  W \\
\hline
\end{tabular}
\end{table*}

\begin{table*}
\caption{\label{tab.rvetaAql} Radial velocities
for \object{$\eta$~Aql}
tabulated with heliocentric Julian date, phase and telescope identifier
as in Tab.\ref{tab.rvSUCas}}
\small
\begin{tabular}{c c r r c| c c r r c| c c r r c}
\hline\hline
HJD & phase & \multicolumn{1}{c}{$V_{\mbox{\scriptsize rad}}$} &
$\sigma$ & Tel.&
HJD & phase & \multicolumn{1}{c}{$V_{\mbox{\scriptsize rad}}$} &
$\sigma$ & Tel.&
HJD & phase & \multicolumn{1}{c}{$V_{\mbox{\scriptsize rad}}$} &
$\sigma$ & Tel.\\
$-2400000$ & & & & &
$-2400000$ & & & & &
$-2400000$ & & & & \\
days & & \multicolumn{1}{c}{\kms} & \multicolumn{1}{c}{\kms} & &
days & & \multicolumn{1}{c}{\kms} & \multicolumn{1}{c}{\kms} & &
days & & \multicolumn{1}{c}{\kms}  & \multicolumn{1}{c}{\kms} & \\
\hline
46253.8931 & 0.981 & $ -29.6$ & 0.5 &  M & 46598.8233 & 0.043 & $ -31.6$ & 0.4 &  T & 46638.7967 & 0.613 & $  -5.1$ & 0.3 &  T \\
46567.9001 & 0.735 & $   5.7$ & 0.5 &  T & 46600.8805 & 0.330 & $ -19.4$ & 0.4 &  T & 46692.7422 & 0.130 & $ -27.8$ & 0.4 &  T \\
46569.9183 & 0.016 & $ -31.4$ & 0.5 &  T & 46601.8224 & 0.461 & $ -15.1$ & 0.4 &  T & 46692.7463 & 0.130 & $ -29.1$ & 0.4 &  T \\
46577.9625 & 0.137 & $ -27.9$ & 0.4 &  T & 46601.8940 & 0.471 & $ -15.0$ & 0.4 &  T & 46693.6991 & 0.263 & $ -21.3$ & 0.3 &  T \\
46578.9693 & 0.277 & $ -20.7$ & 0.4 &  T & 46626.8249 & 0.945 & $ -27.3$ & 0.9 &  T & 46693.7013 & 0.263 & $ -23.3$ & 0.4 &  T \\
46595.8356 & 0.627 & $  -3.6$ & 0.5 &  T & 46626.8277 & 0.945 & $ -28.3$ & 0.7 &  T & 46694.6349 & 0.394 & $ -14.1$ & 0.3 &  T \\
46596.8629 & 0.770 & $   8.3$ & 0.5 &  T & 46628.8916 & 0.233 & $ -23.5$ & 0.9 &  T & 46694.6383 & 0.394 & $ -16.9$ & 0.3 &  T \\
46596.9143 & 0.777 & $   8.6$ & 0.5 &  T & 46628.8938 & 0.233 & $ -25.0$ & 0.8 &  T & 46721.5852 & 0.149 & $ -28.3$ & 0.3 &  T \\
46597.8967 & 0.914 & $ -21.4$ & 0.5 &  T & 46638.7932 & 0.613 & $  -7.3$ & 0.3 &  T &  &  &  &  &  \\
\hline
\end{tabular}
\end{table*}

\begin{table*}
\caption{\label{tab.rvXCyg} Radial velocities
for \object{X~Cyg}
tabulated with heliocentric Julian date, phase and telescope identifier
as in Tab.\ref{tab.rvSUCas}}
\small
\begin{tabular}{c c r r c| c c r r c| c c r r c}
\hline\hline
HJD & phase & \multicolumn{1}{c}{$V_{\mbox{\scriptsize rad}}$} &
$\sigma$ & Tel.&
HJD & phase & \multicolumn{1}{c}{$V_{\mbox{\scriptsize rad}}$} &
$\sigma$ & Tel.&
HJD & phase & \multicolumn{1}{c}{$V_{\mbox{\scriptsize rad}}$} &
$\sigma$ & Tel.\\
$-2400000$ & & & & &
$-2400000$ & & & & &
$-2400000$ & & & & \\
days & & \multicolumn{1}{c}{\kms} & \multicolumn{1}{c}{\kms} & &
days & & \multicolumn{1}{c}{\kms} & \multicolumn{1}{c}{\kms} & &
days & & \multicolumn{1}{c}{\kms}  & \multicolumn{1}{c}{\kms} & \\
\hline
46595.8575 & 0.767 & $  34.7$ & 0.8 &  T & 46699.5437 & 0.095 & $ -18.7$ & 0.6 &  W & 46778.4393 & 0.909 & $  12.0$ & 0.9 &  W \\
46596.8763 & 0.829 & $  19.2$ & 1.1 &  T & 46700.6732 & 0.163 & $ -12.9$ & 0.5 &  W & 46787.5361 & 0.464 & $  13.9$ & 0.6 &  W \\
46596.9284 & 0.832 & $  17.9$ & 1.0 &  T & 46701.4797 & 0.213 & $  -8.4$ & 0.5 &  W & 46804.4272 & 0.495 & $  16.0$ & 0.5 &  W \\
46597.9083 & 0.892 & $  14.6$ & 0.8 &  T & 46702.6193 & 0.282 & $  -2.9$ & 0.7 &  W & 46820.4997 & 0.476 & $  14.1$ & 0.6 &  W \\
46600.8948 & 0.074 & $ -20.7$ & 0.6 &  T & 46712.5208 & 0.886 & $  14.9$ & 0.8 &  W & 46821.4405 & 0.533 & $  20.0$ & 0.5 &  W \\
46601.8446 & 0.132 & $ -15.7$ & 0.5 &  T & 46714.4798 & 0.006 & $ -21.1$ & 0.8 &  W & 46832.4521 & 0.205 & $  -9.8$ & 0.5 &  W \\
46601.9250 & 0.137 & $ -15.4$ & 0.5 &  T & 46715.4700 & 0.066 & $ -21.0$ & 0.7 &  W & 46837.4628 & 0.511 & $  18.3$ & 0.7 &  W \\
46626.9121 & 0.662 & $  29.4$ & 0.8 &  T & 46718.7283 & 0.265 & $  -4.8$ & 0.5 &  W & 46838.4579 & 0.572 & $  23.2$ & 0.7 &  W \\
46626.9145 & 0.662 & $  31.8$ & 0.7 &  T & 46719.5049 & 0.313 & $  -0.1$ & 0.4 &  W & 46841.4611 & 0.755 & $  35.6$ & 0.9 &  W \\
46628.8734 & 0.782 & $  32.7$ & 1.6 &  T & 46721.5968 & 0.440 & $  11.7$ & 0.5 &  T & 46907.8592 & 0.807 & $  24.0$ & 1.1 &  W \\
46628.8757 & 0.782 & $  30.3$ & 0.9 &  T & 46722.6269 & 0.503 & $  16.8$ & 0.5 &  W & 46923.8464 & 0.783 & $  30.5$ & 0.9 &  W \\
46630.8251 & 0.901 & $  15.5$ & 0.9 &  T & 46723.5177 & 0.558 & $  21.9$ & 0.5 &  W & 46929.8353 & 0.148 & $ -14.5$ & 0.5 &  W \\
46630.8282 & 0.901 & $  12.8$ & 0.6 &  T & 46724.6175 & 0.625 & $  27.7$ & 0.6 &  W & 46948.7378 & 0.302 & $  -0.5$ & 0.5 &  W \\
46631.8741 & 0.965 & $  -8.1$ & 1.1 &  T & 46726.4931 & 0.739 & $  35.1$ & 0.6 &  W & 46966.7858 & 0.403 & $   8.5$ & 0.5 &  W \\
46631.8765 & 0.965 & $  -9.0$ & 1.0 &  T & 46727.5534 & 0.804 & $  24.7$ & 1.1 &  W & 46982.7153 & 0.375 & $   6.4$ & 0.5 &  W \\
46638.8183 & 0.389 & $   8.5$ & 0.4 &  T & 46736.5008 & 0.350 & $   4.8$ & 0.5 &  W & 47015.6619 & 0.386 & $   6.6$ & 0.5 &  W \\
46638.8223 & 0.389 & $   6.9$ & 0.4 &  T & 46737.4772 & 0.409 & $   8.8$ & 0.6 &  W & 47020.5889 & 0.687 & $  32.7$ & 0.6 &  W \\
46690.5014 & 0.543 & $  21.6$ & 0.6 &  W & 46744.5603 & 0.842 & $  16.9$ & 1.1 &  W & 47020.6540 & 0.691 & $  33.1$ & 0.6 &  W \\
46691.5741 & 0.608 & $  26.7$ & 0.5 &  W & 46745.4394 & 0.895 & $  13.9$ & 0.8 &  W & 47022.8284 & 0.823 & $  20.0$ & 1.2 &  W \\
46692.7654 & 0.681 & $  34.3$ & 0.6 &  T & 46747.5446 & 0.024 & $ -23.1$ & 0.8 &  W & 47074.5549 & 0.980 & $ -16.6$ & 1.1 &  W \\
46692.7690 & 0.681 & $  32.7$ & 0.6 &  T & 46748.4471 & 0.079 & $ -20.7$ & 0.7 &  W & 47221.9276 & 0.974 & $ -13.3$ & 1.1 &  W \\
46693.7229 & 0.739 & $  36.4$ & 0.6 &  T & 46749.4390 & 0.139 & $ -15.9$ & 0.6 &  W & 47287.9364 & 0.002 & $ -21.1$ & 0.8 &  T \\
46694.6825 & 0.798 & $  26.5$ & 0.9 &  T & 46754.4320 & 0.444 & $  11.7$ & 0.5 &  W & 47336.8573 & 0.988 & $ -17.3$ & 0.9 &  W \\
46695.7006 & 0.860 & $  15.7$ & 1.5 &  W & 46759.6019 & 0.760 & $  34.4$ & 0.6 &  W & 47368.7213 & 0.932 & $   6.6$ & 0.9 &  W \\
46698.7404 & 0.045 & $ -22.4$ & 0.7 &  W & 46760.4299 & 0.810 & $  22.6$ & 1.0 &  W & 47401.7821 & 0.950 & $  -2.1$ & 1.1 &  W \\
\hline
\end{tabular}
\end{table*}

\begin{table*}
\caption{\label{tab.rvTMon} Radial velocities
for \object{T~Mon}
tabulated with heliocentric Julian date, phase and telescope identifier
as in Tab.\ref{tab.rvSUCas}}
\small
\begin{tabular}{c c r r c| c c r r c| c c r r c}
\hline\hline
HJD & phase & \multicolumn{1}{c}{$V_{\mbox{\scriptsize rad}}$} &
$\sigma$ & Tel.&
HJD & phase & \multicolumn{1}{c}{$V_{\mbox{\scriptsize rad}}$} &
$\sigma$ & Tel.&
HJD & phase & \multicolumn{1}{c}{$V_{\mbox{\scriptsize rad}}$} &
$\sigma$ & Tel.\\
days & & \multicolumn{1}{c}{\kms} & \multicolumn{1}{c}{\kms} & &
days & & \multicolumn{1}{c}{\kms} & \multicolumn{1}{c}{\kms} & &
days & & \multicolumn{1}{c}{\kms}  & \multicolumn{1}{c}{\kms} & \\
\hline
46690.8721 & 0.526 & $  31.8$ & 0.7 &  W & 46816.6123 & 0.177 & $   3.7$ & 0.5 &  W & 47185.6968 & 0.829 & $  41.0$ & 0.9 &  W \\
46691.8946 & 0.564 & $  34.4$ & 0.6 &  W & 46819.6325 & 0.288 & $  12.3$ & 0.6 &  W & 47201.5613 & 0.415 & $  22.3$ & 0.7 &  W \\
46693.0200 & 0.605 & $  37.9$ & 0.6 &  T & 46819.7940 & 0.294 & $  12.2$ & 0.6 &  W & 47219.5660 & 0.081 & $  -2.2$ & 0.7 &  W \\
46698.8515 & 0.821 & $  40.3$ & 0.9 &  W & 46820.6233 & 0.325 & $  15.4$ & 0.5 &  W & 47459.9015 & 0.971 & $  -5.9$ & 0.8 &  W \\
46699.8290 & 0.857 & $  40.1$ & 0.9 &  W & 46821.5707 & 0.360 & $  17.9$ & 0.6 &  W & 47460.8602 & 0.006 & $  -6.4$ & 0.8 &  W \\
46701.8752 & 0.933 & $   0.5$ & 1.3 &  W & 46837.4965 & 0.949 & $  -5.1$ & 1.0 &  W & 47461.9020 & 0.045 & $  -4.9$ & 0.7 &  W \\
46712.9006 & 0.341 & $  16.2$ & 0.6 &  W & 46838.4928 & 0.986 & $  -7.6$ & 0.8 &  W & 47462.8425 & 0.080 & $  -2.8$ & 0.6 &  W \\
46714.8649 & 0.413 & $  22.2$ & 0.6 &  W & 46839.5447 & 0.025 & $  -6.5$ & 0.8 &  W & 47463.8720 & 0.118 & $  -0.4$ & 0.6 &  W \\
46718.8628 & 0.561 & $  34.6$ & 0.7 &  W & 46842.6611 & 0.140 & $   1.3$ & 0.5 &  W & 47465.8016 & 0.189 & $   5.0$ & 0.6 &  W \\
46721.9968 & 0.677 & $  42.2$ & 0.7 &  T & 46843.6551 & 0.177 & $   4.5$ & 0.5 &  W & 47468.7922 & 0.300 & $  13.6$ & 0.6 &  W \\
46724.8698 & 0.783 & $  39.7$ & 0.9 &  W & 46844.5609 & 0.211 & $   6.3$ & 0.5 &  W & 47483.8815 & 0.858 & $  39.5$ & 0.9 &  W \\
46737.9196 & 0.266 & $  10.0$ & 0.6 &  W & 46845.5234 & 0.246 & $   9.7$ & 0.6 &  W & 47484.6698 & 0.887 & $  31.8$ & 1.4 &  W \\
46747.8243 & 0.632 & $  39.4$ & 0.7 &  W & 46848.6041 & 0.360 & $  17.7$ & 0.6 &  W & 47487.8679 & 0.005 & $  -7.6$ & 0.8 &  W \\
46748.8349 & 0.670 & $  41.7$ & 0.7 &  W & 46849.4812 & 0.393 & $  21.8$ & 0.6 &  W & 47488.8276 & 0.041 & $  -6.0$ & 0.7 &  W \\
46749.8550 & 0.708 & $  43.1$ & 0.7 &  W & 46869.4885 & 0.133 & $   0.3$ & 0.6 &  W & 47511.6547 & 0.885 & $  32.7$ & 1.2 &  W \\
46754.9007 & 0.894 & $  28.7$ & 1.6 &  W & 46871.5348 & 0.208 & $   6.3$ & 0.5 &  W & 47512.5908 & 0.920 & $   5.3$ & 1.7 &  W \\
46756.8182 & 0.965 & $  -6.4$ & 0.8 &  W & 46878.5509 & 0.468 & $  26.6$ & 0.6 &  W & 47514.6537 & 0.996 & $  -7.6$ & 0.8 &  W \\
46757.9339 & 0.006 & $  -6.8$ & 0.7 &  W & 46879.5312 & 0.504 & $  30.3$ & 0.6 &  W & 47515.6171 & 0.032 & $  -5.8$ & 0.7 &  W \\
46777.7859 & 0.741 & $  41.6$ & 0.8 &  W & 46896.5495 & 0.133 & $   0.9$ & 0.5 &  W & 47518.6512 & 0.144 & $   1.4$ & 0.6 &  W \\
46778.7110 & 0.775 & $  40.7$ & 0.9 &  W & 46897.5349 & 0.170 & $   4.2$ & 0.6 &  W & 47540.6771 & 0.959 & $  -6.1$ & 0.9 &  W \\
46785.7814 & 0.036 & $  -5.3$ & 0.8 &  W & 46908.5168 & 0.576 & $  35.4$ & 0.6 &  W & 47542.6802 & 0.033 & $  -5.8$ & 0.6 &  W \\
46786.7197 & 0.071 & $  -3.5$ & 0.7 &  W & 47104.8231 & 0.837 & $  40.9$ & 0.8 &  W & 47543.6669 & 0.069 & $  -3.9$ & 0.6 &  W \\
46787.7846 & 0.110 & $  -0.8$ & 0.6 &  W & 47105.7735 & 0.872 & $  37.3$ & 1.1 &  W & 47546.6029 & 0.178 & $   3.7$ & 0.5 &  W \\
46788.7740 & 0.147 & $   1.8$ & 0.6 &  W & 47105.8959 & 0.877 & $  36.4$ & 1.1 &  W & 47567.4688 & 0.950 & $  -4.8$ & 1.0 &  W \\
46800.7951 & 0.592 & $  36.3$ & 0.6 &  W & 47106.8466 & 0.912 & $  10.8$ & 2.0 &  W & 47570.6161 & 0.066 & $  -3.4$ & 0.6 &  W \\
46803.6614 & 0.698 & $  42.8$ & 0.7 &  W & 47107.7508 & 0.945 & $  -3.2$ & 1.1 &  W & 47574.5915 & 0.213 & $   6.6$ & 0.5 &  W \\
46804.7016 & 0.736 & $  42.2$ & 0.7 &  W & 47132.8199 & 0.873 & $  38.1$ & 1.1 &  W & 47598.5275 & 0.098 & $  -2.0$ & 0.5 &  W \\
46807.5951 & 0.843 & $  40.6$ & 0.8 &  W & 47160.7206 & 0.905 & $  21.3$ & 2.2 &  W & 47628.5022 & 0.207 & $   6.2$ & 0.5 &  W \\
46809.6504 & 0.919 & $   4.3$ & 1.7 &  W & 47166.7606 & 0.128 & $  -0.4$ & 0.6 &  W &  &  &  &  &  \\
46812.6328 & 0.030 & $  -6.0$ & 0.6 &  W & 47184.7271 & 0.793 & $  40.8$ & 0.9 &  W &  &  &  &  &  \\
\hline
\end{tabular}
\end{table*}

\begin{table*}
\caption{\label{tab.rvRSPup} Radial velocities
for \object{RS~Pup}
tabulated with heliocentric Julian date, phase and telescope identifier
as in Tab.\ref{tab.rvSUCas}}
\small
\begin{tabular}{c c r r c| c c r r c| c c r r c}
\hline\hline
HJD & phase & \multicolumn{1}{c}{$V_{\mbox{\scriptsize rad}}$} &
$\sigma$ & Tel.&
HJD & phase & \multicolumn{1}{c}{$V_{\mbox{\scriptsize rad}}$} &
$\sigma$ & Tel.&
HJD & phase & \multicolumn{1}{c}{$V_{\mbox{\scriptsize rad}}$} &
$\sigma$ & Tel.\\
$-2400000$ & & & & &
$-2400000$ & & & & &
$-2400000$ & & & & \\
days & & \multicolumn{1}{c}{\kms} & \multicolumn{1}{c}{\kms} & &
days & & \multicolumn{1}{c}{\kms} & \multicolumn{1}{c}{\kms} & &
days & & \multicolumn{1}{c}{\kms}  & \multicolumn{1}{c}{\kms} & \\
\hline
46567.6224 & 0.584 & $  39.3$ & 0.7 &  T & 47137.9956 & 0.347 & $  24.5$ & 0.6 &  T & 47226.6620 & 0.487 & $  32.5$ & 0.6 &  T \\
46569.6208 & 0.632 & $  41.4$ & 0.9 &  T & 47139.9518 & 0.395 & $  27.7$ & 0.5 &  T & 47229.6587 & 0.559 & $  37.3$ & 0.6 &  T \\
46816.9178 & 0.599 & $  41.7$ & 0.7 &  T & 47140.9092 & 0.418 & $  29.6$ & 0.6 &  T & 47230.6465 & 0.583 & $  38.3$ & 0.6 &  T \\
46817.8361 & 0.622 & $  41.8$ & 0.7 &  T & 47158.8912 & 0.852 & $  40.6$ & 1.3 &  T & 47250.6180 & 0.065 & $   3.9$ & 0.5 &  T \\
46844.7528 & 0.271 & $  18.8$ & 0.5 &  T & 47159.8937 & 0.876 & $  19.4$ & 2.5 &  T & 47251.7167 & 0.092 & $   5.8$ & 0.5 &  T \\
46865.7343 & 0.777 & $  46.9$ & 0.7 &  T & 47163.9114 & 0.973 & $   0.4$ & 0.8 &  T & 47252.6533 & 0.114 & $   6.9$ & 0.5 &  T \\
46866.7706 & 0.802 & $  48.3$ & 0.7 &  T & 47191.7599 & 0.645 & $  40.8$ & 0.8 &  T & 47253.6804 & 0.139 & $   9.3$ & 0.5 &  T \\
46891.6576 & 0.403 & $  28.3$ & 0.5 &  T & 47192.8026 & 0.670 & $  42.2$ & 0.6 &  T & 47254.6975 & 0.164 & $  10.8$ & 0.5 &  T \\
46892.6733 & 0.427 & $  29.5$ & 0.5 &  T & 47216.7219 & 0.247 & $  17.0$ & 0.5 &  T & 47255.6428 & 0.186 & $  12.8$ & 0.4 &  T \\
46894.6520 & 0.475 & $  32.5$ & 0.6 &  T & 47218.7441 & 0.296 & $  20.5$ & 0.6 &  T & 47281.6106 & 0.813 & $  48.5$ & 0.7 &  T \\
46895.6430 & 0.499 & $  34.2$ & 0.6 &  T & 47219.7635 & 0.321 & $  21.8$ & 0.5 &  T & 47282.6333 & 0.838 & $  46.4$ & 0.9 &  T \\
47108.0423 & 0.625 & $  40.0$ & 0.6 &  T & 47220.7620 & 0.345 & $  23.8$ & 0.5 &  T & 47283.6608 & 0.862 & $  34.9$ & 1.6 &  T \\
47109.9968 & 0.672 & $  42.3$ & 0.7 &  T & 47221.7628 & 0.369 & $  26.6$ & 0.9 &  T & 47285.6202 & 0.910 & $   2.2$ & 1.3 &  T \\
47135.9998 & 0.299 & $  20.9$ & 0.5 &  T & 47224.6869 & 0.439 & $  29.6$ & 0.5 &  T &  &  &  &  &  \\
47136.9995 & 0.323 & $  22.7$ & 0.5 &  T & 47225.6855 & 0.463 & $  32.0$ & 0.6 &  T &  &  &  &  &  \\
\hline
\end{tabular}
\end{table*}
\begin{table*}
\caption{\label{tab.rvSVVul} Radial velocities
for \object{SV~Vul}
tabulated with heliocentric Julian date, phase and telescope identifier
as in Tab.\ref{tab.rvSUCas}}
\small
\begin{tabular}{c c r r c| c c r r c| c c r r c}
\hline\hline
HJD & phase & \multicolumn{1}{c}{$V_{\mbox{\scriptsize rad}}$} &
$\sigma$ & Tel.&
HJD & phase & \multicolumn{1}{c}{$V_{\mbox{\scriptsize rad}}$} &
$\sigma$ & Tel.&
HJD & phase & \multicolumn{1}{c}{$V_{\mbox{\scriptsize rad}}$} &
$\sigma$ & Tel.\\
$-2400000$ & & & & &
$-2400000$ & & & & &
$-2400000$ & & & & \\
days & & \multicolumn{1}{c}{\kms} & \multicolumn{1}{c}{\kms} & &
days & & \multicolumn{1}{c}{\kms} & \multicolumn{1}{c}{\kms} & &
days & & \multicolumn{1}{c}{\kms}  & \multicolumn{1}{c}{\kms} & \\
\hline
46567.9059 & 0.301 & $ -11.1$ & 0.5 &  T & 46701.4865 & 0.266 & $ -14.5$ & 0.6 &  W & 46929.8293 & 0.334 & $  -8.4$ & 0.5 &  W \\
46569.9237 & 0.346 & $  -8.1$ & 0.6 &  T & 46712.4960 & 0.510 & $   3.6$ & 0.6 &  W & 46930.7197 & 0.354 & $  -6.3$ & 0.6 &  W \\
46571.9283 & 0.390 & $  -4.2$ & 0.6 &  T & 46714.4659 & 0.554 & $   6.6$ & 0.6 &  W & 46936.7726 & 0.488 & $   3.2$ & 0.7 &  W \\
46577.8713 & 0.522 & $   5.7$ & 0.6 &  T & 46715.4626 & 0.576 & $   8.6$ & 0.6 &  W & 46948.7335 & 0.754 & $  18.8$ & 0.9 &  W \\
46578.9309 & 0.545 & $   7.5$ & 0.7 &  T & 46719.4919 & 0.665 & $  13.8$ & 0.7 &  W & 46952.7110 & 0.842 & $  21.7$ & 0.9 &  W \\
46595.7931 & 0.920 & $  15.5$ & 1.2 &  T & 46721.5909 & 0.712 & $  16.0$ & 0.8 &  T & 46957.7416 & 0.954 & $  -4.2$ & 1.5 &  W \\
46596.8693 & 0.944 & $   5.6$ & 1.5 &  T & 46722.6102 & 0.735 & $  17.1$ & 0.8 &  W & 46962.7796 & 0.066 & $ -25.7$ & 0.7 &  W \\
46596.9224 & 0.945 & $   4.6$ & 1.5 &  T & 46723.5067 & 0.755 & $  18.4$ & 0.8 &  W & 46963.7815 & 0.088 & $ -25.2$ & 0.6 &  W \\
46597.9034 & 0.967 & $  -6.6$ & 1.4 &  T & 46724.4540 & 0.776 & $  18.3$ & 0.8 &  W & 46965.7643 & 0.132 & $ -23.1$ & 0.6 &  W \\
46600.8878 & 0.033 & $ -24.1$ & 0.8 &  T & 46726.4851 & 0.821 & $  20.4$ & 0.8 &  W & 46966.7971 & 0.155 & $ -21.5$ & 0.6 &  W \\
46601.8328 & 0.054 & $ -25.0$ & 0.6 &  T & 46727.5424 & 0.844 & $  20.0$ & 1.0 &  W & 46986.6386 & 0.595 & $  10.0$ & 0.7 &  W \\
46601.9034 & 0.055 & $ -24.8$ & 0.7 &  T & 46728.4855 & 0.865 & $  19.6$ & 0.9 &  W & 47014.5363 & 0.214 & $ -16.8$ & 0.6 &  W \\
46626.9039 & 0.610 & $  10.6$ & 0.7 &  T & 46737.4858 & 0.065 & $ -25.5$ & 0.8 &  W & 47020.5822 & 0.349 & $  -6.7$ & 0.6 &  W \\
46628.8635 & 0.654 & $  14.3$ & 0.8 &  T & 46739.6430 & 0.113 & $ -25.3$ & 0.6 &  W & 47022.5484 & 0.392 & $  -3.9$ & 0.6 &  W \\
46630.8124 & 0.697 & $  15.7$ & 0.7 &  T & 46747.4292 & 0.286 & $ -12.8$ & 0.5 &  W & 47085.4681 & 0.789 & $  20.4$ & 0.8 &  W \\
46635.8082 & 0.808 & $  19.8$ & 0.8 &  W & 46748.4399 & 0.308 & $ -11.6$ & 0.5 &  W & 47136.4842 & 0.921 & $   9.8$ & 1.6 &  W \\
46638.8090 & 0.875 & $  19.9$ & 0.7 &  T & 46749.4313 & 0.330 & $  -9.8$ & 0.5 &  W & 47225.9357 & 0.907 & $  14.6$ & 0.8 &  W \\
46690.4936 & 0.022 & $ -23.4$ & 0.9 &  W & 46751.5414 & 0.377 & $  -7.0$ & 0.6 &  W & 47226.9296 & 0.929 & $   6.1$ & 1.5 &  W \\
46691.5656 & 0.046 & $ -25.3$ & 0.8 &  W & 46754.4367 & 0.441 & $  -1.2$ & 0.5 &  W & 47227.9149 & 0.951 & $  -7.2$ & 1.4 &  W \\
46692.7545 & 0.072 & $ -24.8$ & 0.6 &  T & 46778.4308 & 0.974 & $ -10.3$ & 1.5 &  W & 47231.8922 & 0.039 & $ -25.7$ & 0.7 &  W \\
46692.7582 & 0.072 & $ -26.0$ & 0.6 &  T & 46780.4389 & 0.018 & $ -22.9$ & 1.1 &  W & 47310.8002 & 0.790 & $  19.3$ & 0.9 &  W \\
46693.7111 & 0.093 & $ -24.9$ & 0.5 &  T & 46787.5283 & 0.176 & $ -21.4$ & 0.6 &  W & 47317.8479 & 0.947 & $  -6.8$ & 1.5 &  W \\
46694.6467 & 0.114 & $ -24.4$ & 0.5 &  T & 46842.9554 & 0.406 & $  -2.3$ & 0.6 &  W & 47400.6032 & 0.784 & $  21.6$ & 0.8 &  W \\
46699.5372 & 0.222 & $ -17.7$ & 0.5 &  W & 46923.8367 & 0.201 & $ -18.6$ & 0.6 &  W & 47404.8114 & 0.877 & $  19.4$ & 1.1 &  W \\
46700.6669 & 0.248 & $ -15.5$ & 0.6 &  W & 46927.7622 & 0.288 & $ -11.9$ & 0.6 &  W & 47407.5810 & 0.938 & $  -5.0$ & 1.4 &  W \\
\hline
\end{tabular}
\end{table*}



\end{document}